%% file: icdm17.tex
\documentclass[conference]{IEEEtran}
\usepackage{graphicx}
\usepackage{balance}
\usepackage{graphicx}
\usepackage{balance}  
\usepackage{booktabs}
\usepackage{amsmath}
\usepackage{mathtools}
\usepackage{amsthm}
\usepackage{amssymb}
\usepackage[linesnumbered,ruled,vlined]{algorithm2e}
\usepackage{ifthen}
\usepackage[overload]{textcase}

\usepackage{graphicx}
\usepackage{subfig}
\usepackage{caption}
\usepackage{booktabs}
\usepackage{multirow}
\usepackage{epstopdf}
\usepackage{colortbl}
\usepackage{balance}
\usepackage{algpseudocode}
\usepackage{enumitem}
\usepackage{varwidth}
\usepackage{multirow}
\usepackage{url}
\usepackage{bm}

\newtheorem{Lemma}{Lemma}
\newtheorem{Definition}{Definition}

\newcommand\I{\mathbb{I}}
\newcommand\E{\mathbb{E}}

\newcommand\R{\mathcal{R}}
\newcommand\Q{\mathcal{Q}}
\newcommand{\Var}{\mathrm{Var}}
\newcommand{\Cov}{\mathrm{C}}
\usepackage{xcolor}

\newlength{\protowidth}

\def\IM{\textsf{IM}}
\def\IE{\textsf{IE}}

\def\PIMA{\textsf{PIMA}}
\def\OPT{\textsf{OPT}}

\def\RIS{\textsf{RIS}}

\def\ROS{\textsf{IIS}}

\def\PIMA{\textsf{SSA}}
\def\DPIMA{\textsf{DSSA}}

\def\G{\mathcal{G}}

\def\SKIM{\textsf{SKIM}}

\def\IMM{\textsf{IMM}}
\def\SKLRIS{\textsf{SKIS}}
\def\SKRIS{\textsf{RIS}}
\def\src{\textsf{src}}
\def\Gain{\textsf{gain}}

\usepackage{siunitx}
\ExplSyntaxOn
\cs_new_eq:NN \calc \fp_eval:n
\ExplSyntaxOff

\def\cca#1{\cellcolor{red!#1}#1}

\DeclareMathOperator*{\argmax}{arg\,max}

\newcommand\EatDot[1]{}

\ifCLASSINFOpdf
\else
\fi
\begin{document}
%
\title{Importance Sketching of Influence Dynamics in Billion-scale Networks}

\author{\IEEEauthorblockN{Hung T. Nguyen, Tri P. Nguyen}
\IEEEauthorblockA{Virginia Commonwealth University\\
	Richmond, VA 23284, USA
	\\ Email: \{hungnt,trinpm\}@vcu.edu}
\and
\IEEEauthorblockN{NhatHai Phan}
\IEEEauthorblockA{New Jersey Institute of Technology\\ 
	Newark, NJ 07104, USA
	\\ Email: phan@njit.edu}
\and
\IEEEauthorblockN{Thang N. Dinh}
\IEEEauthorblockA{Virginia Commonwealth University\\
	Richmond, VA 23284, USA
	\\ Email: tndinh@vcu.edu}}


%


\maketitle

\begin{abstract}
The blooming availability of traces for social, biological, and communication networks opens up unprecedented opportunities in analyzing diffusion processes in networks. However, the sheer sizes of the nowadays networks raise serious challenges in computational efficiency and scalability.

In this paper, we propose a new hyper-graph sketching framework for influence dynamics in networks. The central of our sketching framework, called \SKLRIS, is an efficient importance sampling algorithm that returns only \emph{non-singular} reverse cascades in the network. Comparing to previously developed sketches like \RIS{} and \SKIM{}, our sketch significantly enhances estimation quality while substantially reducing processing time and memory-footprint. Further, we present general strategies of using \SKLRIS{} to enhance existing algorithms for \emph{influence estimation} and \emph{influence maximization} which are motivated by practical applications like viral marketing. Using \SKLRIS{}, we design high-quality influence oracle for seed sets with average estimation error up to 10x times smaller than those using \RIS{} and 6x times smaller than \SKIM{}s. In addition, our influence maximization using \SKLRIS{} substantially improves the quality of solutions for greedy algorithms. It achieves up to 10x times speed-up and 4x memory reduction for the fastest \RIS-based DSSA algorithm, while maintaining the same theoretical guarantees. 
\end{abstract}


%
\IEEEpeerreviewmaketitle

\newcommand*{\Scale}[2][4]{\scalebox{#1}{$#2$}}%
\newcommand*{\Resize}[2]{\resizebox{#1}{!}{$#2$}}%
\input{body/introduction}
\input{body/model}
\input{body/fpras}
\input{body/infmax}
\input{body/experiment}
\section{Conclusion}
\label{sec:con}
We propose \SKLRIS{} - a novel sketching tools to approximate influence dynamics in the networks. We provide both comprehensive theoretical and empirical analysis to demonstrate the superiority in size-quality trade-off of \SKLRIS{} in comparisons to the existing sketches. The application of \SKLRIS{} to existing algorithms on Influence Maximization leads to significant performance boost and easily scale to billion-scale networks. In future, we plan to extend \SKLRIS{} to other settings including evolving networks and time-based influence dynamics.

\bibliographystyle{IEEEtran}
\bibliography{social,pids,targetedIM,budgetedIM,infEst,sketch}

\appendix
\input{body/appendix}
\end{document}

%% file: body/introduction.tex
\section{Introduction}
\label{sec:intro}

Online social networks (OSNs) such as Facebook and Twiter have connected billions of users, providing gigantic communication platforms for exchanging and disseminating information. For example, Facebook now has nearly 2 billions monthly active users and more than 2.5 billion pieces of content exchanged daily. Through OSNs, companies and participants have actively capitalized on the ``word-of-mouth'' effect to trigger viral spread of various kinds of information, including marketing messages, propaganda, and even fake news. In the past decade,  a great amount of research has focused on analyzing how information and users' influence propagate within the network, e.g.,  evaluating influence of a group of individuals, aka \emph{influence estimation} \cite{Lucier15,Cohen14,Ohsaka16}, and finding a small group of influential individuals, aka \emph{influence maximization} \cite{Kempe03,Cohen14,Borgs14, Tang14,Tang15,Nguyen163}, or controlling diffusion processes via structure manipulation \cite{Tong12}.

Yet, diffusion analysis is challenging due to the sheer size of the networks. For example, state-of-the-art solutions for influence maximization, e.g., DSSA \cite{Nguyen163}, IMM \cite{Tang15}, TIM/TIM+\cite{Tang14}, cannot complete in the networks with only few million edges \cite{Arora17}. Further, our comprehensive experiments on influence estimation show that the average estimation error can be as high as 40-70\% for popular sketches such as \RIS{} and \SKIM{}. This happens even on a small network with only 75K nodes (Epinion). This calls for development of new approaches for analyzing influence dynamics in large-scale networks.

In this paper, we propose a new importance sketching technique, termed \SKLRIS{}, that consists of \emph{non-singular} reverse influence cascades, or simply non-singular cascades. Each non-singular cascade simulates the reverse diffusion process from a source node. It is important that each non-singular cascade must include at least another node other than the source itself. Thus, our sketch, specifically, suppresses \emph{singular} cascades that die prematurely at the source. Those singular cascades,  consisting of 30\%-80\% portion in the previous sketches \cite{Cohen14, Borgs14}, not only waste the memory space and processing time but also reduce estimation efficiency of the sketches.  Consequently, \SKLRIS{} contains samples of smaller variances providing estimations of high concentration with less memory and running time. Our new sketch also powers a new principle and scalable influence maximization class of methods, that inherits the algorithmic designs of existing algorithms on top of \SKLRIS{} sketch. Particularly, \SKLRIS{}-based \IM{} methods are the only provably good and efficient enough that can scale to networks of billions of edges across different settings. We summarize of our contributions as follows:



\begin{itemize}[noitemsep,nolistsep]
	\item At the central of our sketch is an importance sampling algorithm to sample non-singular cascades (Alg.~\ref{alg:eris}). For simplicity, we first present the sketch and its sampling algorithm using the popular \emph{independent cascade} model \cite{Kempe03}, and later extend them to other diffusion models.

	\item 	 We provide general frameworks to apply \SKLRIS{} for existing algorithms for the influence estimation and influence maximization problems. We provide theoretical analysis to show that using \SKLRIS{} leads to improved influence estimation oracle due to smaller sample variances and better concentration bounds; and that the state-of-the-art methods for influence maximization like D-SSA \cite{Nguyen163}, IMM \cite{Tang15}, and, TIM/TIM+\cite{Tang14} can also immediately benefit from our new sketch.
	
	\item We conduct comprehensive empirical experiments to demonstrate the effectiveness of our sketch in terms of quality, memory and computational time.  Using SKIS, we can design high-quality influence oracle for seed set with average estimation error up to 10x times smaller than those using \RIS{} and 6x times those using \SKIM{}. In addition, our influence maximization using \SKLRIS{} substantially improves the quality of solutions for greedy algorithms. It achieves up to 10x times speed-up and 4x memory reduction for the fastest \RIS-based DSSA algorithm, while maintaining the same theoretical guarantees. 
\end{itemize}

\textbf{Related work.}
Sketching methods have become extremely useful for dealing with problems in massive sizes. Most notable sketches including bottom-$k$ sketches \cite{Cohen07} a summary of a set of items with nonnegative weights, count-min sketch \cite{Cormode05count} that count the frequency of different events in the stream.

Recently, Cohen et al. \cite{Cohen14} investigate the combined reachability sketch which is a bottom-$k$ min-hash sketch of the set of reachable nodes. They show small estimation errors for estimating influences. However, this method deteriorates for large influences since the size of each sketch is fixed. Similar scheme was applied for continuous-time model \cite{Du13}.

Borgs et al. \cite{Borgs14} proposed reverse influence sketch (\RIS) which captures the influences in a reverse manner. This approach has advantage in estimating large influences and becomes very successful in finding the seed set with maximum influence, i.e. influence maximization. \cite{Ohsaka16} uses \RIS{} sketches to estimate influences in dynamic graphs. Other related works on influences in multiplex networks and identifying the sources of influence cascades are studied in \cite{Nguyen13_l,Shen12,Nguyen164}.

\textbf{Organization.} The rest of the paper is organized as follows: In Section~\ref{sec:model}, we introduce the diffusion model and two problems of influence estimation/maximization. We propose our importance sketching scheme in Section~\ref{sec:sketching}. Applications in influence estimation/maximization is presented in Sections~\ref{sec:oracle} and \ref{sec:max}, respectively. Extensions to other diffusion models are discussed in Section~\ref{sec:ext} which is followed by experiments in Section~\ref{sec:exps} and conclusion in Section~\ref{sec:con}.

%% file: body/model.tex
\section{Preliminaries}
\label{sec:model}
Consider a social network abstracted as a  graph $\G = (V, E, w)$.  Each edge $(u,v) \in E$ is associated with a real number $w(u,v) \in [0, 1]$ specifying the probability that node $u$ will influence $v$ once $u$ is influenced.
To model the influence dynamic in the network, we first focus on the popular \textit{Independent Cascade} (IC) model \cite{Kempe03} and then, discuss the extensions of our techniques to other models, e.g. Linear Threshold (LT) or Continuous-time model, later in Section~\ref{sec:ext}.
%

\subsection{Independent Cascade Model}

For a subset of nodes $S \subseteq V$, called seed set, the influence propagation from $S$ happens in discrete rounds $t=0, 1,...$ At round $0$, only nodes in $S$ are active (aka influenced) and the others are inactive. Each newly activated node $u$ at round $t$ will have a single chance to activate each neighbor $v$ of $u$ with probability $w(u, v)$. An activated node remains active till the end of the diffusion propagation. The process stops when no more nodes get activated.

{\bf Sample Graphs.} Once a node $u$ gets activated, it will activate each of its neighbor $v$ with probability $w(u,v)$. This can be thought of as flipping a biased coin that gives head with probability $w(u, v)$ to determine whether the edge $(u, v)$ exists. If the coin lands head for the edge $(u,v)$, the activation occurs and we call $(u,v)$ a \textit{live-edge}. Since all the influences in the IC model are independent, it does not matter when coins are flipped to determine the states of the edges. Thus, we can flip all the coins at the beginning instead of waiting until $u$ gets activated. We call the deterministic graph $g$ that contains all the live-edges resulted from a series of coin flips over all the edges in $\G$ a \textit{sample graph} of $\mathcal G$.

\textbf{Probabilistic Space.} The set of all sample graphs generated from $\mathcal G$ together with their probabilities define a probabilistic space $\Omega_\G$. Each sample graph $g \in \Omega_\G$ can be generated by flipping coins on all the edges to determine whether or not the edge is live or appears in $g$. That is each edge $(u, v)$ will be present in a sample graph with probability $w(u, v)$. Therefore, a sample graph $g=(V, E' \subseteq E)$ is generated from $\G$ with a probability $\Pr[g \sim \G]$ calculated by,
\begin{align}
	\Pr[g \sim \G] = \prod_{(u,v) \in E'} w(u,v) \prod_{(u,v) \notin E'} (1-w(u,v)).
\end{align}

{\bf Influence Spread.} Given a diffusion model, the measure \emph{Influence Spread} (or simply \emph{influence}) of a seed set $S$ is defined as the expected number of active nodes in the end of the diffusion propagation, where the expectation is taken over the probabilistic space $\Omega_\G$. Given a sample graph $g \sim \G$ and a seed set $S \subset V$, we denote $\eta_g(S)$ the set of nodes reachable from $S$ (including nodes in $S$ themselves). The influence spread of $S$ is defined as follows,
\begin{align}
	\I(S) = \sum_{g \sim \G} |\eta_g(S)| \Pr[g \sim \G].
\end{align}

The frequently used notations are summarized in Table~\ref{tab:syms}.
\renewcommand{\arraystretch}{1.2}

\setlength\tabcolsep{3pt}
\begin{table}[hbt]\small
	\centering
	\caption{Table of notations}
	\begin{tabular}{p{1.5cm}|p{6.5cm}}
		\addlinespace
		\toprule
		\bf Notation  &  \quad \quad \quad \bf Description \\
		\midrule 
		$n, m$ & \#nodes, \#edges of graph $\G=(V, E, w)$.\\
		\hline
		$\I(S), \hat \I(S)$ & Expected Influence of $S \subseteq V$ and an estimate.\\
		\hline
		$N^{in}(S)$ & Set of in-neighbor nodes of $S$.\\
		\hline
		$\gamma_v,\Gamma$ & $\gamma_v = 1-\Pi_{u \in N^{in}(v)}(1-w(u,v)); \Gamma = \sum_{v \in V} \gamma_v$.\\
		\hline
		$\gamma_0$ & $\gamma_0 = \sum_{v \in V}\gamma_v/n$.\\
		\hline
		$R_j, \R$ & A random \ROS{} sample and a \SKLRIS{} sketch.\\
		\hline
		$\Cov_\R(S)$ & $\Cov_{\R}(S) = |R_j \in \R | \R_j \cap S \neq \emptyset|$.\\
		\bottomrule
	\end{tabular}%
	\label{tab:syms}%
	\vspace{-0.1in}
\end{table}%

\subsection{Influence Estimation/Maximization Problems}
We describe the tasks of Influence Estimation and Maximization which are used to evaluate sketches' efficiency.

\begin{Definition}[Influence Estimation (\IE{})] Given a probabilistic graph $\G$ and a seed set of nodes $S \subseteq V$, the \IE{} problem asks for an estimation $\hat \I(S)$ of the set influence $\I(S)$.
\end{Definition}

\begin{Definition}[Influence Maximization (\IM{}) \cite{Kempe03}] Given a probabilistic graph $\G$, a budget $k$, the \IM{} problem asks for a set $S_k$ of size at most $k$ having the maximum influence among all other sets of size at most $k$,
	\begin{equation}
		S_k = \argmax_{S \subseteq V, |S| \leq k} \I(S).
	\end{equation}	
\end{Definition}

%
%

\subsection{Sketch-based Methods for IE/IM}

\subsubsection{Reverse Influence Sketch (RIS)}

Essentially, a random \RIS{} sample, denoted by $R_j$, contains a \textit{random} set of nodes, following a diffusion model, that can influence a \textit{randomly selected} source node, denoted by $\src(R_j)$. A \RIS{} sample is generated in three steps:
\begin{itemize}[noitemsep,nolistsep]
	\item[1)] Select a random node $v \in V$ which serves as $\src(R_j)$.
	\item[2)] Generate a sample graph $g \sim \mathcal{G}$.
	\item[3)] Return the set $R_j$ of nodes that can reach $v$ in $g$.
\end{itemize}

Thus, the probability of generating a particular \RIS{} sample $R_j$ can be computed based on the source selection and the sample graphs that has $R_j$ as the set of nodes that reach $\src(R_j)$ in $g$. Let denote such set of nodes that can reach to a node $v$ in sample graph $g$ by $\eta^{-}_g(v)$. We have,
\begin{align}
\label{eq:ris_prob}
\Pr[R_j] = \frac{1}{n} \sum_{g, \eta^-_g(\src(R_j)) = R_j} \Pr[g].
\end{align}

The key property of \RIS{} samples for influence estimation/maximization is stated in the following lemma.
\begin{Lemma}[\cite{Borgs14}]
	\label{lam:borgs14}
	Given a random \RIS{} sample  $R_j$ generated from $\mathcal{G} = (V,E,w)$, for a set $S \subseteq V$ of nodes, we have,
	\begin{align}
	\I(S) = n \cdot \Pr[R_j \cap S \neq \emptyset].
	\end{align}
\end{Lemma}
Thus, estimating/maximizing $\I(S)$ is equivalent to estimating/maximizing the probability $\Pr[R_j \cap S \neq \emptyset]$.

\textbf{Using \RIS{} samples for \IE{}/\IM{}.} Thanks to Lemma~\ref{lam:borgs14}, a general strategy for \IE{}/\IM{} is generating a set of \RIS{} samples, then returning an empirical estimate of $\Pr[R_j \cap S \neq \emptyset]$ on generated samples for \IE{} or the set $\hat S_k$ that intersects with most samples for \IM{}. The strong advantage of \RIS{} is the reuse of samples to estimate influence of any seed set $S \subseteq V$. Ohsaka et al. \cite{Ohsaka16} build a query system to answer influence queries. \cite{Borgs14,Tang14,Tang15,Nguyen163,Nguyen173} recently use \RIS{} samples in solving Influence Maximization problem with great successes, i.e. handling large networks with tens of millions of nodes and billions of edges.

\subsubsection{Combined Reachability Sketch (SKIM)}

Cohen et al. \cite{Cohen14} proposed the combined reachability sketch which can be used to estimate influences of multiple seed sets. Each node $u$ in the network is assigned a combined reachability sketch which is a bottom-$k$ min-hash sketch of the set of nodes reachable from $u$ in $l$ sample graphs. \cite{Cohen14} generates $l$ sample graphs $g$ of $\G$, i.e. $l = 64$ by default, and build a combined reachability sketch of size $k$ for each node. 

The influence estimate of a seed set $S$ is computed by taking the bottom-$k$ sketch of the union over all the sketches of nodes in $S$ and applying the cardinality estimator \cite{Cohen14}. Using the sketches, the solution for \IM{} is found by following the greedy algorithm which repeatedly adds a node with highest marginal influence into the solution. Here, the marginal influences are similarly estimated from node sketches.

\begin{figure}[!t]
	\vspace{-0.1in}
	\centering
	\subfloat[Weighted Cascade (\textsf{WC})]{
		\includegraphics[width=0.4\linewidth]{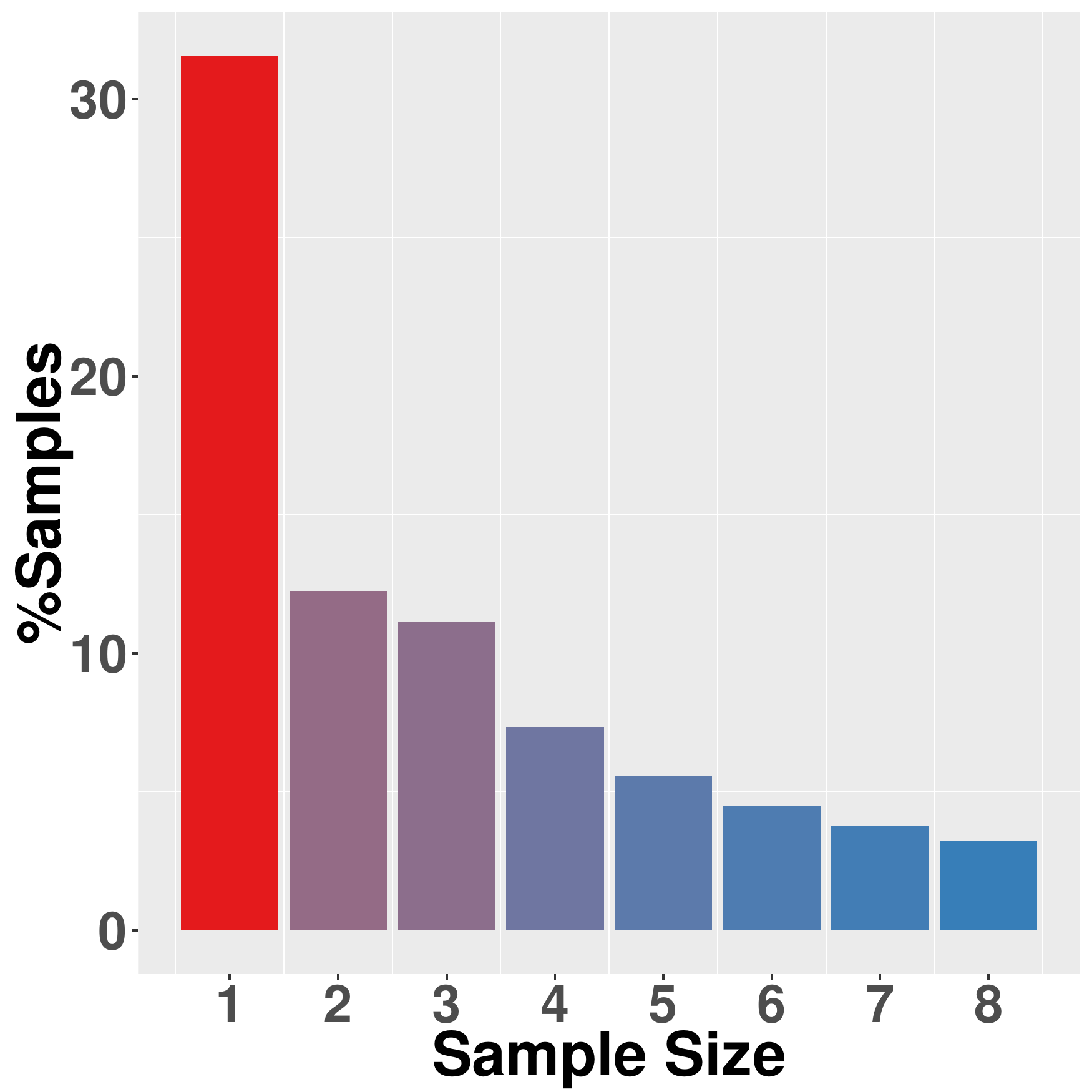}
	} 
	\subfloat[Trivalency (\textsf{TRI})]{
		\includegraphics[width=0.4\linewidth]{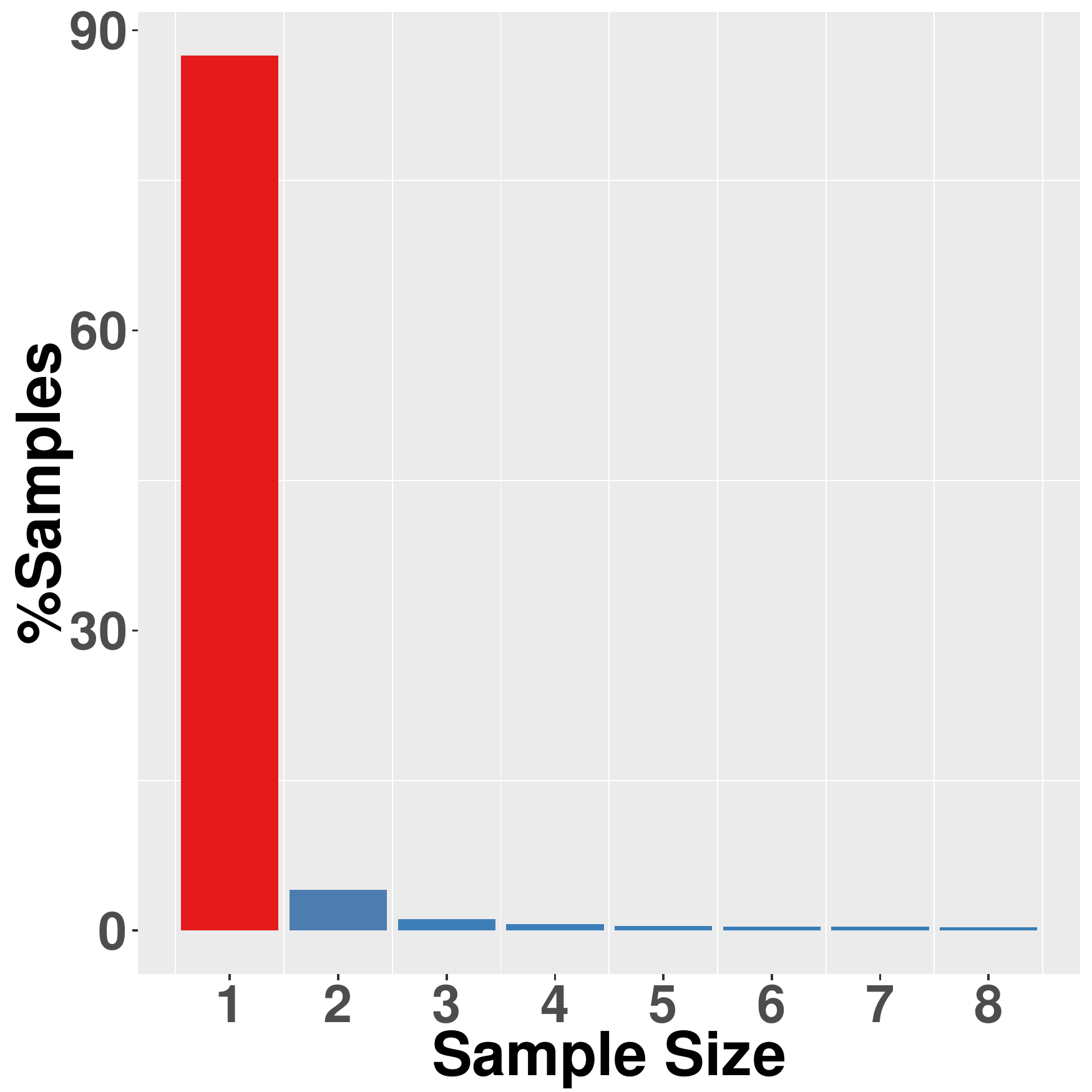}
	}
	\caption{Distribution of reversed cascade sizes on Epinions network (and on other networks as well) with two different edge weight models: Weighted Cascade (\textsf{WC}) and Trivalency (\textsf{TRI}). The majority of the cascades are singular.}
	\label{fig:size_dist}
	\vspace{-0.25in}
\end{figure}

\textbf{Common Shortcomings.} According to recent benchmarks \cite{Arora17} and our own empirical evaluations (details in Section~\ref{sec:exps}), both \RIS{} and \SKIM{} yield significant influence estimation errors. For \RIS{}, it is due to the fact that the majority of \RIS{} samples contain only their sources as demonstrated in Figure~\ref{fig:size_dist} with up to 86\% of such \RIS{} samples overall. These samples, termed \textit{singular}, harm the performance in two ways: 1) they do not contribute to the influence computation of other seed sets than the ones that contain the sources, however, the contribution is known in advance, i.e. number of seed nodes; 2) these samples magnify the variance of the \RIS{}-based random variables used in estimation causing high errors.

This motivates our Importance Sketching techniques to generate only \textit{non-singular} samples that are useful for influence estimations of many seed sets (not just ones with the sources).

%% file: body/fpras.tex
\section{Importance Sketching}
\label{sec:sketching}

This section introduces our core construction of \textit{Importance Sketching} algorithm to generate random \textit{non-singular} samples with probabilities proportional to those in the original sample space of reverse influence samples and normalized by the probability of generating a non-singular ones.

\begin{algorithm} \small
	\caption{Importance Influence Sampling (\ROS{}) Alg.}
	\label{alg:eris}
	\KwIn{Graph $\G = (V,E,w)$}
	\KwOut{$R_j$ - A random \ROS{} sample}
	Pick a node $v \in V$ as the source with probability in Eq.~\ref{eq:probu};\\
	Select an in-neighbor $u_i$ of $v$, $u_i \in N^{in}(v)$, with probability of selecting $u_i$ given in Eq.~\ref{eq:probn};\\
	Initialize a queue $Q = \{u_i\}$ and a node set $R_j = \{v,u_i\}$;\\
	\ForEach{$u_t \in N^{in}(v), t \neq i$}{
		With probability $w(u_t, v)$: \\
		\qquad $Q.\textsf{push}(u_t); R_j \leftarrow R_j \cup \{u_t\};$
	}
	\While{$Q$ is not empty}{
		$v = Q.\textsf{pop}()$;\tcp{get the longest inserted node}
		\ForEach{$u \in N^{in}(v)\backslash (R_j \cup Q)$}{
			With probability $w(u,v)$:\\
			\qquad $Q.\textsf{push}(u); R_j \leftarrow R_j \cup \{u\}$; \tcp{insert $u$}
		}
	}
	\textbf{return} $R_j$;
\end{algorithm}
\vspace{-0.05in}

\subsection{Importance Influence Sampling (\ROS{})}
\textbf{Sample Spaces and Desired Property.} Let $\Omega_{\RIS}$ be the sampling space of reverse influence samples (\RIS{}) with probability $\Pr[R_j \in \Omega_{\RIS}]$ of generating sample $R_j$. Let $\Omega_{\SKLRIS}$ be a subspace of $\Omega_{\RIS}$ and corresponds to the space of \textit{only} non-singular reverse influence samples in $\Omega_{\RIS}$. Since $\Omega_{\SKLRIS}$ is a subspace of $\Omega_{\RIS}$, the probability $\Pr[R_j \in \Omega_{\SKLRIS}]$ of generating a non-singular sample from $\Omega_{\SKLRIS}$ is larger than that from $\Omega_{\RIS}$. Specifically, for a node $v \in V$, let $\gamma_v$ be the probability of generating a non-singular sample if $v$ is selected as the source and $\Gamma = \sum_{v \in V} \gamma_v$. Then, since the sample sources are selected randomly, the ratio of generating a non-singular sample to generating any sample in $\Omega_{\RIS}$ is $\frac{\Gamma}{n}$ and thus, the probability $\Pr[R_j \in \Omega_{\SKLRIS}]$ is as follows,
\begin{align}
	\Pr[R_j \in \Omega_{\SKLRIS}] = \frac{n}{\Gamma}\Pr[R_j \in \Omega_{\RIS}].
\end{align}
Our upcoming \ROS{} algorithm aims to achieve this desired property of sampling non-singular samples from $\Omega_{\SKLRIS}$.


\textbf{Sampling Algorithm.} Our Importance Influence Sampling (\ROS{}) scheme involves three core components:

\begin{itemize}
	\item[1)] \textit{Probability of having a non-singular sample.} For a node $v \in V$, a sample with source $v$ is singular if no in-neighbor of $v$ is selected, that happens with probability $\Pi_{u \in N^{in}(v)}(1-w(u,v))$. Hence, the probability of having a non-singular sample from a node $v$ is the complement:
	\begin{align}
		\label{eq:gamma_v}
		\gamma_v = 1-\Pi_{u \in N^{in}(v)}(1-w(u,v)).
	\end{align}
	\item[2)] \textit{Source Sampling Rate.} Note that the set of non-singular samples is just a subset of all possible samples and we want to generate uniformly random samples from that subset. Moreover, each node $v$ has a probability $\gamma_v$ of generating a non-singular sample from it. Thus, in order to generate a random sample, we select $v$ as the source with probability $\Pr[\src(R_j) = v]$ computed as follows,
	\begin{align}
		\label{eq:probu}
		\Pr[\src(R_j) = v] = \frac{\gamma_v}{\sum_{u \in V} \gamma_u} = \frac{\gamma_v}{\Gamma},
	\end{align}
	where $\Gamma = \sum_{u \in V} \gamma_u$, and then generate a uniformly random non-singular sample from the specific source $v$ as described in the next component.
	\item[3)] \textit{Sample a non-singular sample from a source.} From the $\src(R_j) = v$, we generate a non-singular sample $R_j$ from $v$ uniformly at random. Let $N^{in}(v) = \{ u_1, u_2, \dots, u_l \}$ be a fixed-order set of in-neighbors of $v$. We divide the all possible non-singular samples from $v$ into $l$ buckets: bucket $B_i, 1 \leq i \leq l$ contains those samples that have the first node from $N^{in}(v)$ being $u_i$. That means all the nodes $u_1, \dots, u_{i-1}$ are not in the sample but $u_i$ is in for certain. The other nodes from $u_{i+1}$ to $u_l$ may appear and will be sampled following the normal \RIS{} sampling. Now we select the bucket that $R_j$ belongs to with the probability of selecting $B_i$ being as follows,
	\begin{align}
		\label{eq:probn}
		\Pr[\text{select }B_i]=\frac{\prod_{t = 1}^{i-1}(1-w(u_t,v)) w(u_i,v)}{\gamma_v}.
	\end{align}
	
	\noindent For $i = 1$, we have $\Pr[\text{select } B_1] = w(u_1,v)$. Note that $\sum_{i = 1}^l \Pr[\text{select } B_i] = 1$. Assume bucket $B_i$ is selected and, thus, node $u_i$ is added as the second node besides the source into $R_j$. For each other node $u_t, t \neq i$, $u_t$ is selected into $R_j$ with probability $w(u_t, v)$ following the ordinary \RIS{} for the IC model.
\end{itemize}

These three components guarantee a non-singular sample. The detailed description of \ROS{} sampling is in Alg.~\ref{alg:eris}. The first step selects the source of the \ROS{} sample among $V$. Then, the first incoming node to the source $v$ is picked (Line~2) following the above description of the component 3). Each of the other incoming neighbors also tries to influence the source (Lines~4-6). The rest performs similarly as in \RIS{} \cite{Borgs14}. That is for each newly selected node, its incoming neighbors are randomly added into the sample with probabilities equal to their edge weights. It continues until no newly selected node is observed. Note that Line~3 only adds the selected neighbors $u_i$ of $v$ into $Q$ but adds both $v$ and $u_i$ to $R_j$. The loop from Lines~7-11 mimics the BFS-like sampling procedure of \RIS{}.

Let $\Pr[R_j]$ be the probability of generating a non-singular sample $R_j$ using \ROS{} algorithm. We have
\begin{align}
	\Pr[R_j] & = \sum_{v \in V}\Pr[\src(R_j) = v] \Pr[\text{generate } R_j \text{ from }v] \nonumber \\
	& = \sum_{v \in V}\frac{\gamma_v}{\Gamma} \frac{\Pr[R_j \in \Omega_{\RIS} \text{ and } \src(R_j) = v]}{\gamma_v} \nonumber \\
	& = \frac{n}{\Gamma} \sum_{v \in V} \frac{1}{n} \Pr[R_j \in \Omega_{\RIS} \text{ and } \src(R_j) = v] \nonumber \\
	& = \frac{n}{\Gamma} \Pr[R_j \in \Omega_{\RIS}] = \Pr[R_j \in \Omega_{\SKLRIS}], \nonumber
\end{align}

\noindent where $\Pr[\text{generate } R_j \text{ from }v] = \frac{\Pr[R_j \in \Omega_{\RIS} \text{ and } \src(R_j) = v]}{\gamma_v}$ due to the selection of the bucket that $R_j$ belongs to in \ROS{}. Thus, the output $R_j$ of \ROS{} is an random sample from non-singular space $\Omega_{\SKLRIS}$ and we obtain the following lemma.
\begin{Lemma}
	Recall that $\Omega_{\SKLRIS{}}$ is the sample space of non-singular reverse influence samples. \ROS{} algorithm generates a random non-singular sample from sample space $\Omega_{\SKLRIS{}}$.
\end{Lemma}

\textbf{Connection between IIS Samples and Influences.}
We establish the following key lemma that connects our \ROS{} samples with the influence of any seed set $S$.
\begin{Lemma}
	Given a random \ROS{} sample $R_j$ generated by Alg.~\ref{alg:eris} from the graph $\G=(V,E,w)$, for any set $S \subseteq V$, we have,
	\begin{align}
		\label{eq:lem6}
		\I(S) = \Pr[R_j \cap S \neq \emptyset] \cdot \Gamma + \sum_{v \in S} (1-\gamma_v),
	\end{align}
	
	\noindent where $\gamma_v$ and $\Gamma$ are defined in Eqs.~\ref{eq:gamma_v} and \ref{eq:probu}.
	\label{lem:multi_set_rois}
\end{Lemma}
The proof is presented in our extended version \cite{extended}. The influence $\I(S)$ of any set $S$ comprises of two parts: 1) $\Pr[R_j \cap S \neq \emptyset] \cdot \Gamma$ depends on the randomness of $R_j$ and 2) the fixed amount $\sum_{v \in S} (1-\gamma_v)$ that is inherent to set $S$ and accounts for the contribution of singular samples in $\Omega_{\RIS}$ to the influence $\I(S)$. Lemma~\ref{lem:multi_set_rois} states that instead of computing or estimating the influence $\I(S)$ directly, we can equivalently compute or estimate $\Pr[R_j \cap S \neq \emptyset] \cdot \Gamma + \sum_{v \in S} (1-\gamma_v)$ using \ROS{} samples.

\textbf{Remark:} Notice that we can further generate samples of larger sizes and reduce the variance as shown later, however, the computation would increase significantly.

\section{Influence Oracle via IIS Sketch (SKIS)}
\label{sec:oracle}
We use \ROS{} sampling to generate a sketch for answering influence estimation queries of different node sets. We show that the random variables associated with our samples have much smaller variances than that of \RIS{}, and hence, lead to better concentration or faster estimation with much fewer samples required to achieve the same or better quality.

\textbf{\SKLRIS{}-based Influence Oracle.}
An \SKLRIS{} sketch $\R$ is a collection of \ROS{} samples generated by Alg.~\ref{alg:eris}, i.e. $\R = \{R_1, \dots, R_T\}$. As shown in Lemma~\ref{lem:multi_set_rois}, the influence $\I(S)$ can be estimated through estimating the probability $\Pr[R_j \cap S \neq \emptyset]$.
Thus, from a \SKLRIS{} sketch $\R = \{R_1, \dots, R_T\}$, we can obtain an estimate $\hat \I(S)$ of $\I(S)$ for any set $S$ by,
\begin{align}
	\label{eq:eq21}
	\hat \I_\R(S)  = \frac{\Cov_\R(S)}{|\R|}\cdot \Gamma + \sum_{v \in S} (1-\gamma_v),
\end{align}
where $\Cov_\R(S)$ is coverage of $S$ on $\R$, i.e.,
\begin{align}
	\Cov_\R(S) = |\{R_j \in \R | R_j \cap S \neq \emptyset\}|.
\end{align}

\begin{algorithm} \small
	\caption{\SKLRIS{}-based Influence Oracle}
	\label{alg:oracle}
	\KwIn{Graph $\G = (V,E,w)$}
	Preprocessing: Generate a \textsf{SKIS} sketch $\R = \{R_1, \dots, R_T\}$ of \ROS{} samples using Alg.~\ref{alg:eris}.\\
	For any influence query for any set $S$: return $\hat \I_\R(S)$ (Eq.~\ref{eq:eq21}).
\end{algorithm}

We build an \SKLRIS{}-based oracle for influence queries by generating a set $\R$ of $T$ \ROS{} samples in a preprocessing step and then answer influence estimation query $\hat \I_\R(S)$ for any requested set $S$ (Alg.~\ref{alg:oracle}). In the following, we show the better estimation quality of our sketch through analyzing the variances and estimating concentration properties.

\textbf{\SKLRIS{} Random Variables for Estimations.}
For a random \ROS{} sample $R_j$ and a set $S$, we define random variables:
\newcommand{\twopartdef}[3]
{
	\left\{
	\begin{array}{ll}
		#1 & \mbox{if } #2 \\
		#3 & \mbox{otherwise.}
	\end{array}
	\right.
}
\begin{align}
\label{eq:varx}
X_j(S) = \twopartdef {1} {R_j \cap S \neq \emptyset} {0}, \text{and}\\
\label{eq:varzs}
Z_j(S) = \frac{X_j(S) \cdot \Gamma + \sum_{v \in S}(1-\gamma_v)}{n}.
\end{align}
Then, the means of $X_j(S)$ and $Z_j(S)$ are as follows,
\begin{align}
\label{eq:xvar}
&\E[X_j(S)] = \Pr[R_j \cap S \neq \emptyset] = \frac{\I(S)-\sum_{v\in S}(1-\gamma_v)}{\Gamma} \\
\label{eq:zvar}
&\E[Z_j(S)] = \E[X_j(S)] \cdot \frac{\Gamma}{n} + \frac{\sum_{v \in S} (1-\gamma_v)}{n} = \frac{\I(S)}{n}.
\end{align}


Hence, we can construct a corresponding set of random variables $Z_1(S), Z_2(S), \dots, Z_T(S)$ by Eqs.~\ref{eq:varx} and~\ref{eq:varzs}. Then, $\hat \I_\R(S)=\frac{n}{T}\sum_{j=1}^{T}Z_j(S)$ is an empirical estimate of $\I(S)$ based on the \SKLRIS{} sketch $\R$.

For comparison purposes, let $Y_j(S)$ be the random variable associated with \RIS{} sample $Q_j$ in a \SKRIS{} sketch $\Q$,
\begin{align}
	Y_j(S) = \twopartdef {1} {Q_j \cap S \neq \emptyset} {0}
\end{align}
From Lemma~\ref{lam:borgs14}, the mean value of $Y_j(S)$ is then,
\begin{align}
	\E[Y_j(S)] = \frac{\I(S)}{n}.
\end{align}

\textbf{Variance Reduction Analysis. } We show that the variance of $Z_j(S)$ for \SKLRIS{} is much smaller than that of $Y_j(S)$ for \RIS{}. The variance of $Z_j(S)$ is stated in the following.

\begin{Lemma}
	\label{lem:zvar}
	The random variable $Z_j(S)$ (Eq.~\ref{eq:varzs}) has
	\begin{align}
		\label{eq:eq20}
		\Var&[Z_j(S)] = \frac{\I(S)}{n}\frac{\Gamma}{n} - \frac{\I^2(S)}{n^2} \nonumber \\
		& - \frac{\sum_{v \in S}(1-\gamma_v)}{n^2} (\Gamma + \sum_{v \in S}(1-\gamma_v) - 2 \I(S)).
	\end{align}
\end{Lemma}

Since the random variables $Y_j(S)$ for \RIS{} samples are Bernoulli and $\E[Y_j(S)] = \frac{\I(S)}{n}$, we have $\Var[Y_j(S)] = \frac{\I(S)}{n}(1-\frac{\I(S)}{n})$. Compared with $\Var[Z_j(S)]$, we observe that since $\frac{\Gamma}{n} \leq 1$, $\frac{\I(S)}{n}\frac{\Gamma}{n} - \frac{\I^2(S)}{n^2} \leq \frac{\I(S)}{n} - \frac{\I^2(S)}{n^2} = \Var[Y_j(S)]$,
\begin{align}
	\Var[Z_j(S)]& \leq \Var[Y_j(S)] \nonumber \\
	&- \frac{\sum_{v \in S}(1-\gamma_v)}{n^2} (\Gamma + \sum_{v \in S}(1-\gamma_v) - 2 \I(S)). \nonumber
\end{align}

In practice, most of seed sets have small influences, i.e. $\I(S) \ll \frac{\Gamma}{2}$, thus, $\Gamma + \sum_{v \in S}(1-\gamma_v) - 2 \I(S) \gg 0$. Hence, $\Var[Z_j(S)] < \Var[Y_j(S)]$ holds for most seed sets $S$.

\textbf{Better Concentrations of \SKLRIS{} Random Variables.} Observe that $Z_j(S) \in \Big[\frac{\sum_{v \in S}(1-\gamma_v)}{n}, \frac{\Gamma + \sum_{v \in S}(1-\gamma_v)}{n} \Big]$, we obtain another result on the variance of $Z_j(S)$ as follows.
\begin{Lemma}
	\label{lem:var_z}
	The variance of random variable $Z_j(S)$ satisfies
	\begin{align}
		\Var[Z_j(S)] \leq \frac{\I(S)}{n} \frac{\Gamma}{n}.
	\end{align}
\end{Lemma}
Using the above result with the general form of Chernoff's bound in Lemma~2 in \cite{Tang15}, we derive the following concentration inequalities for random variables $Z_j(S)$ of \SKLRIS{}.
\begin{Lemma}
	\label{lem:bound}
	Given a \SKLRIS{} sketch $\R = \{ R_1, \dots, R_T \}$ with random variables $Z_1(S), \dots, Z_T(S)$, we have,
	\begin{align}
		&\Pr\Big[\frac{\sum_{j = 1}^{T} Z_j(S)}{T}n - \I(S) \geq \epsilon \I(S)\Big] \leq \exp\Big( \frac{-\epsilon^2 T}{2\boldsymbol{\frac{\Gamma}{n}}+\frac{2}{3}\epsilon} \frac{\I(S)}{n} \Big) \nonumber \\
		&\Pr\Big[\frac{\sum_{j = 1}^{T} Z_j(S)}{T}n - \I(S) \leq -\epsilon \I(S)\Big] \leq \exp\Big( \frac{-\epsilon^2 T}{2\boldsymbol{\frac{\Gamma}{n}}} \frac{\I(S)}{n}\Big). \nonumber
	\end{align}
\end{Lemma}
Compared with the bounds for \RIS{} sketch in Corollaries~1 and 2 in \cite{Tang15}, the above concentration bounds for \SKLRIS{} sketch (Lemma~\ref{lem:bound}) are stronger, i.e. tighter. Specifically, we have the factor $\frac{\Gamma}{n}$ in the denominator of the $\exp(.)$ function while for \RIS{} random variables, it is simply $1$.

\textbf{Sufficient Size of \SKLRIS{} Sketch for High-quality Estimations.}
There are multiple strategies to determine the number of \ROS{} samples generated in the preprocessing step. For example, \cite{Ohsaka16} generates samples until total size of all samples reaches $O(\frac{1}{\epsilon^3}(n+m)\log(n))$. Generating \ROS{} samples to reach such a specified threshold is vastly faster than using \RIS{} due to the bigger size of \ROS{} samples. This method provides an \textit{additive} estimation error guarantee within $\epsilon$. Alternatively, by Lemma~\ref{lem:bound}, we derive the sufficient number of \ROS{} samples to provide the more preferable $(\epsilon,\delta)$-estimation of $\I(S)$.
\begin{Lemma}
	\label{lem:monte}
	Given a set $S$, $\epsilon, \delta \geq 0$, if the \SKLRIS{} sketch $\R$ has at least $(2\frac{\Gamma}{n}+\frac{2}{3}\epsilon) \ln(\frac{2}{\delta}) \frac{n}{\I(S)}\epsilon^{-2}$ \ROS{} samples, $\hat \I_\R(S)$ is an $(\epsilon,\delta)$-estimate of $\I(S)$, i.e.,
	\begin{align}
		\Pr[(1-\epsilon)\I(S) \leq \hat \I_\R(S) \leq (1+\epsilon)\I(S)] \geq 1-\delta.
	\end{align}
\end{Lemma}
In practice, $\I(S)$ is unknown in advance and a lower-bound of $\I(S)$, e.g. $|S|$, can be used to compute the necessary number of samples to provide the same guarantee. Compared to \RIS{} with weaker concentration bounds, we save a factor of $O(\frac{\Gamma}{n})$.



%% file: body/infmax.tex
\section{SKIS-based IM Algorithms}
\label{sec:max}


With the help of \SKLRIS{} sketch that is better in estimating the influences compared to the well-known successful \RIS{}, we can largely improve the efficiency of \IM{} algorithms in the broad class of \RIS{}-based methods, i.e. RIS \cite{Borgs14}, TIM/TIM+ \cite{Tang14}, IMM \cite{Tang15}, BCT \cite{Nguyen162,Nguyen172}, SSA/DSSA \cite{Nguyen163}. This improvement is possible since these methods heavily rely on the concentration of influence estimations provided by \RIS{} samples.


\textbf{\SKLRIS{}-based framework.}
Let $\R = \{R_1, R_2, \dots \}$ be a \SKLRIS{} sketch of \ROS{} samples. $\R$ gives an influence estimate
\begin{align}
	\label{eq:compute_inf}
	\hat \I_\R(S) = \hat \E_\R[Z_j(S)] \cdot n = \frac{\Cov_\R(S)}{|\R|} \cdot \Gamma + \sum_{v \in S}(1-\gamma_v),
\end{align}

\noindent for any set $S$. Thus, instead of optimizing over the exact influence, we can intuitively find the set $S$ to maximize the estimate function $\hat \I(S)$. Then, the framework of using \SKLRIS{} sketch to solve \IM{} problem contains two main steps:
\begin{itemize}
	\item[1)] Generate a \SKLRIS{} sketch $\R$ of \ROS{} samples,
	\item[2)] Find the set $S_k$ that maximizes the function $\hat \I_R(S)$ and returning $S_k$ as the solution for the \IM{} instance.
\end{itemize}
There are two essential questions related to the above \SKLRIS{}-based framework
: 1) Given a \SKLRIS{} sketch $\R$ of \ROS{} samples, how to find $S_k$ of $k$ nodes that maximizes $\hat \I_\R(S_k)$ (in Step 2)? 2) How many \ROS{} samples in the \SKLRIS{} sketch $\R$ (in Step 1) are sufficient to guarantee a high-quality
solution for \IM?


We give the answers for the above questions in the following sections. Firstly, we adapt the gold-standard greedy algorithm to obtain an $(1-(1-1/k)^k)$-approximate solution over a \SKLRIS{} sketch. Secondly, we adopt recent techniques on \RIS{} with strong solution guarantees to \SKLRIS{} sketch.


\begin{algorithm} \small
	\caption{\textsf{Greedy} Algorithm on \SKLRIS{} sketch}
	\label{alg:cov}
	\KwIn{\SKLRIS{} sketch $\R$ and $k$}
	\KwOut{An $(1-(1-1/k)^k)$-approximate seed set $\hat S_k$}
	$\hat S_k = \emptyset$\\
	\For{$i=1:k$}{
		$\hat v \leftarrow \arg \max_{v\in V\backslash \hat S_k}\big (\frac{\Delta_\R(v, \hat S_k)}{|\R|}\Gamma + (1-\gamma_v)\big )$\\
		Add $\hat v$ to $\hat S_k$\\
	}
	\textbf{return} $\hat S_k$
\end{algorithm}

\subsection{Greedy Algorithm on \SKLRIS{} Sketches}
Let consider the optimization problem of finding a set $S_k$ of at most $k$ nodes to maximize the function $\hat \I_\R(S)$ on a \SKLRIS{} sketch $\R$ of \ROS{} samples under the cardinality constraint $|S| \leq k$. The function $\hat \I_\R(S)$ is monotone and submodular since it is the weighted sum of a set coverage function $\Cov_\R(S)$ and a linear term $\sum_{v \in S} (1-\gamma_v)$. Thus, we obtain the following lemma with the detailed proof in our extended version \cite{extended}.
\begin{Lemma}
	Given a set of \ROS{} samples $\R$, the set function $\hat \I_\R(S)$ defined in Eq.~\ref{eq:compute_inf} is monotone and submodular.
	\label{lem:sub}
\end{Lemma}
Thus, a standard greedy scheme \cite{Nemhauser81}, which iteratively selects a node with highest marginal gain, gives an $(1-(1-\frac{1}{k})^k)$, that converges to $(1-1/e)$ asymptotically, approximate solution $\hat S_k$. The marginal gain of a node $v$ with respect to a set $S$ on \SKLRIS{} sketch $\R$ is defined as follows,
\begin{align}
	\label{eq:margin}
	\Gain_\R(v,S) = \frac{\Delta_\R(v,\hat S_k)}{|\R|}\Gamma + (1-\gamma_v),
\end{align}
where $\Delta_\R(v,S) = \Cov_\R(S \cup \{v\}) - \Cov_\R(S)$ is called the marginal coverage gain of $v$ w.r.t. $S$ on \SKLRIS{} sketch $\R$.

Given a collection of \ROS{} samples $\R$ and a budget $k$, the Greedy algorithm is presented in Alg.~\ref{alg:cov} with a main loop (Lines~2-4) of $k$ iterations. Each iteration picks a node $\hat v$ having largest marginal gain (Eq.~\ref{eq:margin}) with respect to the current partial solution $\hat S_k$ and adds it to $\hat S_k$.
The approximation guarantee of the \textsf{Greedy} algorithm (Alg.~\ref{alg:cov}) is stated below.
\begin{Lemma}
	\label{lem:greedy}
	The \textsf{Greedy} algorithm (Alg.~\ref{alg:cov}) returns an $(1-(1-\frac{1}{k})^k)$-approximate solution $\hat S_k$,
	\begin{align}
	\hat \I_\R(\hat S_k) \geq (1-(1-\frac{1}{k})^k) \hat \I_\R(S^*_\R),
	\end{align}
	where $S_\R^*$ is the optimal cover set of size $k$ on sketch $\R$.
\end{Lemma}
The lemma is derived directly from the $1-(1-\frac{1}{k})^k$ approximation factor of the ordinary greedy algorithm \cite{Nemhauser81}.

\subsection{Sufficient Size of \SKLRIS{} Sketch for \IM{}}
Since the \SKLRIS{} sketch offers a similar greedy algorithm with approximation ratio $(1-(1-1/k)^k)$ to the traditional \RIS{}, we can combine \SKLRIS{} sketch with any \RIS{}-based algorithm, e.g. RIS\cite{Borgs14}, TIM/TIM+\cite{Tang14}, IMM\cite{Tang15}, BCT\cite{Nguyen172}, SSA/DSSA\cite{Nguyen163}. We discuss the adoptions of two most recent and scalable algorithms, i.e. IMM\cite{Tang15} and SSA/DSSA\cite{Nguyen163}.

\textbf{\IMM+\SKLRIS{}.} Tang et al. \cite{Tang15} provide a theoretical threshold
\begin{align}
	\theta_{\RIS} = O\Big((\log{n \choose k} + \log \delta^{-1})\frac{n}{\OPT_k}\epsilon^{-2}\Big)
\end{align}
on the number of \RIS{} samples to guarantee an $(1-1/e-\epsilon)$-approximate solution for \IM{} problem with probability $1-\delta$. 

Replacing \RIS{} with \ROS{} samples to build a \SKLRIS{} sketch enables us to use the better bounds in Lemma~\ref{lem:bound}. By the approach of \IMM{} in \cite{Tang15} with Lemma~\ref{lem:bound}, we reduce the threshold of samples to provide the same quality to,
%
	\begin{align}
		\theta_{\SKLRIS{}} = O\Big(\frac{\Gamma+k}{n} \theta_{\RIS}\Big).
	\end{align}

\textbf{\PIMA/\DPIMA{}+\SKLRIS{}.} More recently, Nguyen et al. \cite{Nguyen163} propose \PIMA{} and \DPIMA{} algorithms which implement the Stop-and-Stare strategy of alternating between finding candidate solutions and checking the quality of those candidates at exponential points, i.e. $2^{t}, t \geq 1$, to detect a satisfactory solution at the earliest time. 

Combining \SKLRIS{} with \PIMA{} or \DPIMA{} brings about multiple benefits in the checking step of \PIMA{}/\DPIMA{}. The benefits stem from the better concentration bounds which lead to better error estimations and smaller thresholds to terminate the algorithms. The details are in our extended version \cite{extended}.

\section{Extensions to other diffusion models}
\label{sec:ext}
The key step in extending our techniques for other diffusion models is devising an importance sketching procedure for each model. Fortunately, following the same designing principle as \ROS{}, we can devise importance sketching procedures for many other diffusion models. We demonstrate this through two other equally important and widely adopted diffusion models, i.e. Linear Threshold \cite{Kempe03} and Continuous-time model \cite{Du13}.

\textbf{Linear Threshold model \cite{Kempe03}.} This model imposes a constraint that the total weights of incoming edges into any node $v \in V$ is at most 1, i.e. $\sum_{u \in N^{in}(v)} w(u,v) \leq 1$. Every node has a random activation threshold $\lambda_v \in [0,1]$ and gets activated if the total edge weights from active in-neighbors exceeds $\lambda_v$, i.e. $\sum_{u \in N^{in}(v), u \text{ is active}} w(u,v) \geq \lambda_v$. A \RIS{} sampling for LT model \cite{Nguyen172} selects a random node as the source and iteratively picks at most one in-neighbor of the last activated node with probability being the edge weights, $w(u,v)$. 

The importance sketching algorithm for the LT model has the following components:
\begin{itemize}
	\item \textit{Probability of having a non-singular sample}:
	\begin{align}
		\gamma_v = \sum_{u \in N^{in}(v)} w(u,v)
	\end{align}
	\item \textit{Source Sampling Rate}:
	\begin{align}
		\Pr[\src(R_j) = v] = \frac{\gamma_v}{\sum_{v \in V} \gamma_v}
	\end{align}
	\item \textit{Sample a non-singular sample from a source.}: select exactly one in-neighbor $u$ of $\src(R_j)=v$ with probability $\frac{w(u,v)}{\gamma_v}$. The rest follows \RIS{} sampling \cite{Nguyen162}.
\end{itemize}


\textbf{Continuous-time model \cite{Du13}.} 
Here we have a deadline parameter $T$ of the latest activation time and each edge $(u,v)$ is associated with a length distribution, represented by a density function $\mathcal{L}_{(u,v)}(t)$, of how long it takes $u$ to influence $v$. A node $u$ is influenced if the length of the shortest path from any active node at time $0$ is at most $T$. The \RIS{} sampling for the Continuous-time model \cite{Tang15} picks a random node as the source and invokes the Dijkstra's algorithm to select nodes into $\src(R_j)$. When the edge $(u,v)$ is first visited, the activation time is sampled following its length distribution $\mathcal{L}_{(u,v)}(t)$. 
From the length distribution, we can compute the probability $p(u,v,T)$ of an edge $(u,v)$ having activation time at most $T$
\begin{align}
	p(u,v,T) = \int_{t=0}^{T} \mathcal{L}_{(u,v)}(t) dt
\end{align}

The importance sketching procedure for the Continuous-time model has the following components:
\begin{itemize}
	\item \textit{Probability of having a non-singular sample}:
	\begin{align}
		\gamma_v = 1 - \prod_{u \in N^{in}(v)} (1-p(u,v,T))
	\end{align}
	
	\item \textit{Source Sampling Rate}:
	\begin{align}
		\Pr[\src(R_j) = v] = \frac{\gamma_v}{\sum_{v \in V} \gamma_v}
	\end{align}
	
	\item \textit{Sample a non-singular sample from a source.}: Use a bucket system on $p(u,v,T)$ similarly to \ROS{} to select the first in-neighbor $u$. The activation time of $u$ follows the normalized density function $\frac{\mathcal{L}_{(u,v)}(t)}{\gamma_v}$. Subsequently, it continues by following \RIS{} sampling \cite{Tang15}.
\end{itemize}

%% file: body/experiment.tex
\section{Experiments}
\label{sec:exps}
 
We demonstrate the advantages of our \SKLRIS{} sketch through a comprehensive set of experiments on the key influence estimation and maximization problems. Due to space limit, we report the results under the IC model and partial results for the LT model. However, the implementations for all models will be released on our website to produce complete results.
\renewcommand{\arraystretch}{0.9}


\subsection{Experimental Settings}

\setlength\tabcolsep{3pt}
\begin{table}[!h] \small
	\caption{Datasets' Statistics}
	\label{tab:data_sum}
	\centering
	\begin{tabular}{ l  r  r  r}\toprule
		\textbf{Dataset} & \bf \#Nodes& \bf \#Edges & \bf Avg. Degree\\\midrule
		NetPHY & $37\cdot10^3$ & $181\cdot10^3$ & 9.8\\
		Epinions & $75\cdot10^3$ & $841\cdot10^3$ & 22.4\\
		DBLP & $655\cdot10^3$ & $2\cdot10^6$ & 6.1\\
		Orkut & $3\cdot10^6$ & $234\cdot10^6$ & 78.0\\
		Twitter \cite{Kwak10} & $41.7\cdot10^6$ & $1.5\cdot10^9$ & 70.5\\
		Friendster & $65.6\cdot10^6$ & $3.6\cdot10^9$ & 109.6\\
		\bottomrule
		\hline
	\end{tabular}
\end{table}

\renewcommand{\arraystretch}{1}
\textbf{Datasets.} 
We use 6 real-world datasets from \cite{snap,Kwak10} with size ranging from tens of thousands to as large as 65.6 million nodes and 3.6 billion edges. Table \ref{tab:data_sum} gives a summary.

\textbf{Algorithms compared.} On influence estimation, we compare our \textsf{SKIS} sketch with:
\begin{itemize}
	\item \SKRIS{} \cite{Borgs14}: The well-known \RIS{} sketch.
	\item \SKIM{} \cite{Cohen14}: Combined reachability sketch. We run \SKIM{} with default parameters in \cite{Cohen14} ($k = l = 64$). \SKIM{} is modified to read graph from files instead of internally computing the edge weights.
\end{itemize}
Following \cite{Ohsaka16}, we generate samples into \SKLRIS{} and \SKRIS{} until the total size of all the samples reaches $h \cdot n \log n$ where $h$ is a constant. Here, $h$ is chosen in the set $\{5, 10\}$.

On influence maximization, we compare:
\begin{itemize}
	\item \textsf{PMC} \cite{Ohsaka14}: A Monte-Carlo simulation pruned method with no guarantees. It only works on the IC model.
	\item \IMM{} \cite{Tang15}: \RIS{}-based algorithm with quality guarantees.
	\item \DPIMA{} \cite{Nguyen163}: The current fastest \RIS{}-based algorithm with approximation guarantee.
	\item \DPIMA{}+\textsf{SKIS}: A modified version of \DPIMA{} where \SKLRIS{} sketch is adopted to replace \RIS{}.
\end{itemize}
We set $\epsilon = 0.5, \delta = 1/n$ for the last three algorithms. For \textsf{PMC}, we use the default parameter of 200 DAGs.

\textbf{Metrics.}
We compare the  algorithms in terms of the solution quality, running time and memory usage. To compute the solution quality of a seed set, we adopt the relative difference which is defined as $\frac{|\hat \I(S) - \I(S)|}{\max\{\I(S), \hat \I(S)\}} \cdot 100\%$, where $\hat \I(S)$ is an estimate, and $\I(S)$ is the ``ground-truth" influence of $S$. 
%

\textbf{Ground-truth Influence.} Unlike previous studies \cite{Tang15,Cohen14,Ohsaka14} using a constant number of cascade simulations, i.e. 10000, to measure the ground-truth influence with \textit{unknown accuracy}, we adopt the Monte-Carlo Stopping-Rule algorithm \cite{Dagum00} that guarantees an estimation error less than $\epsilon$ with probability at least $1-\delta$ where $\epsilon = 0.005, \delta = 1/n$. Specifically, let $W_j$ be the size of a random influence cascade and $Z_j = \frac{W_j}{n}$ with $\E[Z_j] = \I(S)/n$ and $0 \leq Z_j \leq 1$. The Monte-Carlo method generates sample $Z_j$ until
$\sum_{j = 1}^{T} Z_j \geq 4(e-2)\ln(\frac{2}{\delta})\frac{1}{\epsilon^2}$
and returns $\hat \I(S) = \frac{\sum_{j = 1}^{T} Z_j}{T}n$ as the ground-truth influence.

For Twitter and Friendster dataset, we set $\epsilon = 0.05$, and $\delta = 1/n$ to compute ground-truth due to the huge computational cost in these networks. For the other networks, we keep the default setting of $\epsilon$ and $\delta$ as specified above.

%

\textbf{Weight Settings.}
We consider two widely-used models:
\begin{itemize}
	\item \textit{Weighted Cascade} (\textsf{WC}) \cite{Tang14,Cohen14,Tang15,Nguyen163}: The weight of edge $(u,v)$ is inversely proportional to the in-degree of node $v$, $d_{in}(v)$, i.e. $w(u,v) = \frac{1}{d_{in}(v)}$.
	\item \textit{Trivalency} (\textsf{TRI}) \cite{Cohen14, Chen10, Jung12}: The weight $w(u,v)$ is selected randomly from the set $\{0.1, 0.01, 0.001\}$.
\end{itemize}
\textbf{Environment.} 
We implemented our algorithms in C++ and obtained the implementations of others from the corresponding authors. We conducted all experiments on a CentOS machine with Intel Xeon E5-2650 v3 2.30GHz CPUs and 256GB RAM. We compute the ground-truth for our experiments in a period of 2 months on a cluster of 16 CentOS machines, each with 64 Intel Xeon CPUs X5650 2.67GHz and 256GB RAM.

\renewcommand{\arraystretch}{1}
\setlength\tabcolsep{0.5pt}
\begin{table}[ht] \small
	\vspace{0.05in}
	\centering
	\caption{Average relative differences (\textsf{dnf}: ``did not finish"  within 24h). \SKLRIS{}  almost always returns the lowest errors.}
	\begin{tabular}{llrrrrrrrr|r r rrrrr} 
		\toprule
		& & \multicolumn{7}{c}{\textbf{\textsf{WC} Model}} && \multicolumn{7}{c}{\textbf{\textsf{TRI} Model}}   \\
		\cmidrule{3-9} \cmidrule{11-17}
		& & \multicolumn{2}{c}{\SKLRIS{}} && \multicolumn{2}{c}{\SKRIS{}} && \SKIM{} && \multicolumn{2}{c}{\SKLRIS{}} && \multicolumn{2}{c}{\SKRIS{}} && \SKIM{}  \\
		\cmidrule{3-4} \cmidrule{6-7} \cmidrule{9-9} \cmidrule{11-12} \cmidrule{14-15} \cmidrule{17-17}    
		
		\multirow{1}{*}{{$|S|$}} & \textsf{Nets} & $h(5)$ & $h(10)$ && $h(5)$ & $h(10)$ && $k(64)$ && $h(5)$ & $h(10)$ && $h(5)$ & $h(10)$ && $k(64)$ \\
		
		\midrule		
		\multirow{6}{*}{l}& PHY & \cca{6.2} & \cca{3.7} && \cca{14.0} & \cca{7.8} && \cca{7.5} && \cca{1.7} & \cca{1.3} && \cca{11.8} & \cca{8.2} && \cca{4.5} \\
		& Epin. & \cca{4.7} & \cca{3.0} && \cca{15.7} & \cca{11.8} && \cca{19.6} && \cca{16.6} & \cca{14.2} && \cca{55.3} & \cca{47.4} && \cca{27.7} \\
		& DBLP & \cca{3.8} & \cca{4.1} && \cca{13.7} & \cca{11.6} && \cca{5.0} && \cca{0.9} & \cca{0.7} && \cca{9.4} & \cca{6.4} && \cca{3.5} \\
		& Orkut & \cca{10.3} & \cca{9.2} && \cca{13.5} & \cca{8.8} && \cca{77.6} && \cca{9.3} & \cca{9.9} && \cca{14.5} & \cca{10.8} && \cellcolor{red} \textsf{dnf} \\ 
		& Twit. & \cca{10.9} & \cca{10.5} && \cca{21.4} & \cca{16.0} && \cca{29.1} && \cca{81.4} & \cca{81.9} && \cca{80.8} & \cca{81.5} && \cellcolor{red} \textsf{dnf} \\
		& Frien. & \cca{15.9} & \cca{10.2} && \cca{22.2} & \cca{13.3} && \cellcolor{red} \textsf{dnf} && \cca{29.8} & \cca{21.3} && \cca{28.5} & \cca{23.6} && \cellcolor{red} \textsf{dnf} \\	
		
		\midrule
		\multirow{6}{*}{$10^2$}& PHY & \cca{0.9} & \cca{0.6} && \cca{1.0} & \cca{0.7} && \cca{2.1} && \cca{0.3} & \cca{0.2} && \cca{1.1} & \cca{0.9} && \cca{1.8} \\
		& Epin. & \cca{1.0} & \cca{0.7} && \cca{1.0} & \cca{1.0} && \cca{7.6} && \cca{0.2} & \cca{1.5} && \cca{4.4} & \cca{1.8} && \cca{2.8} \\
		& DBLP & \cca{0.9} & \cca{0.6} && \cca{1.9} & \cca{1.4} && \cca{5.0} && \cca{0.8} & \cca{0.7} && \cca{5.5} & \cca{5.3} && \cca{5.5} \\
		& Orkut & \cca{0.9} & \cca{0.6} && \cca{1.1} & \cca{0.7} && \cca{56.5} && \cca{0.1} & \cca{0.2} && \cca{4.2} & \cca{0.9} && \cellcolor{red} \textsf{dnf} \\
		& Twit. & \cca{1.1} & \cca{1.2} && \cca{1.3} & \cca{1.1} && \cca{60.2} && \cca{4.3} & \cca{3.1} && \cca{6.4} & \cca{5.5} && \cellcolor{red} \textsf{dnf} \\
		& Frien. & \cca{0.9} & \cca{0.7} && \cca{0.9} & \cca{0.7} && \cellcolor{red} \textsf{dnf} && \cca{1.9} & \cca{1.9} && \cca{0.6} & \cca{2.0} && \cellcolor{red} \textsf{dnf} \\
		
		\midrule
		\multirow{6}{*}{$10^3$}{} & PHY & \cca{0.6} & \cca{0.8} && \cca{0.9} & \cca{1.0} && \cca{0.6} && \cca{0.3} & \cca{0.4} && \cca{1.2} & \cca{1.3} && \cca{1.1} \\
		& Epin. & \cca{0.6} & \cca{0.6} && \cca{0.6} & \cca{0.7} && \cca{2.3} && \cca{2.3} & \cca{0.3} && \cca{1.9} & \cca{4.6} && \cca{1.5} \\
		& DBLP & \cca{0.2} & \cca{0.3} && \cca{0.2} & \cca{0.2} && \cca{1.7} && \cca{0.1} & \cca{0.0} && \cca{0.3} & \cca{0.2} && \cca{0.3} \\
		& Orkut & \cca{0.3} & \cca{0.3} && \cca{0.3} & \cca{0.3} && \cca{50.7} && \cca{2.5} & \cca{1.1} && \cca{6.8} & \cca{2.1} && \cellcolor{red} \textsf{dnf} \\
		& Twit. & \cca{0.9} & \cca{0.9} && \cca{1.0} & \cca{0.9} && \cca{36.3} && \cca{0.9} & \cca{2.4} && \cca{4.1} & \cca{2.8} && \cellcolor{red} \textsf{dnf} \\
		& Frien. & \cca{0.3} & \cca{0.3} && \cca{0.3} & \cca{0.2} && \cellcolor{red} \textsf{dnf} && \cca{1.9} & \cca{1.9} && \cca{0.6} & \cca{2.0} && \cellcolor{red} \textsf{dnf} \\ 
		\bottomrule		
	\end{tabular}
	\label{tbl:diff2_wc_all}
\end{table}
\renewcommand{\arraystretch}{1}
\setlength\tabcolsep{0.4pt}
\begin{table}[ht] \small
	\centering
	\caption{Sketch construction time and index memory of algorithms on different edge models. \SKLRIS{} and \SKRIS{} uses roughly the same time and memory and less than that of \SKIM{}.} 
	\begin{tabular}{cl rrrrrrrr | r rrrrrr}
		\toprule
		& &\multicolumn{7}{c}{\textbf{Index Time}} && \multicolumn{7}{c}{\textbf{Index Memory}} \\
		& &\multicolumn{7}{c}{[second (or \textsf{h} for hour)]}  &  & \multicolumn{7}{c}{[MB (or \textsf{G} for GB)]}  \\
		\cmidrule{3-9}\cmidrule{11-17}
		& & \multicolumn{2}{c}{\SKLRIS{}} && \multicolumn{2}{c}{\SKRIS{}} && \SKIM{} && \multicolumn{2}{c}{\SKLRIS{}} && \multicolumn{2}{c}{\SKRIS{}} && \SKIM{} \\
		\cmidrule{3-4} \cmidrule{6-7} \cmidrule{9-9} \cmidrule{11-12} \cmidrule{14-15} \cmidrule{17-17}  
		
		\textbf{M} & \textbf{Nets} & $h(5)$ & $h(10)$ && $h(5)$ & $h(10)$ && $k(64)$ && $h(5)$ & $h(10)$ && $h(5)$ & $h(10)$ && $k(64)$  \\		
		\midrule			
		\multirow{6}{*}{\textsf{WC}} 
		& PHY & 0 & 1 && 1 & 1 && 2  && 41 & 83 && 52 & 105 && 105 \\
		& Epin. & 1 & 1 && 1 & 1 && 10  && 63 & 126 && 81 & 162 && 220 \\
		& DBLP & 10 & 18 && 7 & 14 && 37  && 702 & 1G && 848 & 2G && 2G \\
		& Orkut & 92 & 157 && 69 & 148 && 0.6h && 2G & 5G && 3G & 5G && 9G \\
		& Twit. & 0.6h & 0.9h && 0.4h & 1.0h && 5.2h && 38G & 76G && 42G & 84G && 44G \\			
		& Frien. & 0.8h & 1.8h && 0.8h  & 1.9h  && \textsf{dnf}  && 59G & 117G && 61G  & 117G  && \textsf{dnf} \\	
		\midrule 
		\multirow{6}{*}{\textsf{TRI}} 
		& PHY & 0 & 1 && 1 & 2 && 1 && 46 & 90 && 97 & 194 && 99 \\
		& Epin. & 1 & 1 && 1 & 1 && 29 && 41 & 82 && 41 & 84 && 230 \\
		& DBLP & 11 & 34 && 18 & 36 && 22 && 1G & 2G && 2G & 5G && 2G \\
		& Orkut & 88 & 206 && 89 & 197 && \textsf{dnf}  && 2G & 4G && 2G & 4G && \textsf{dnf} \\
		& Twit. & 0.6h & 1.2h && 0.5h & 1.3h && \textsf{dnf} && 36G & 69G && 36G & 69G && \textsf{dnf} \\
		& Frien. & 0.9h & 2.3h && 1.0h & 2.4h && \textsf{dnf}  && 54G & 108G  &&  54G & 108G && \textsf{dnf} \\
		\bottomrule	
	\end{tabular}
	\label{tbl:sk_cstr_all}
	\vspace{-0.2in}
\end{table}

\begin{figure*}[!ht]
	\centering
	\subfloat[\textsf{TRI} model]{
		\includegraphics[width=0.27\linewidth]{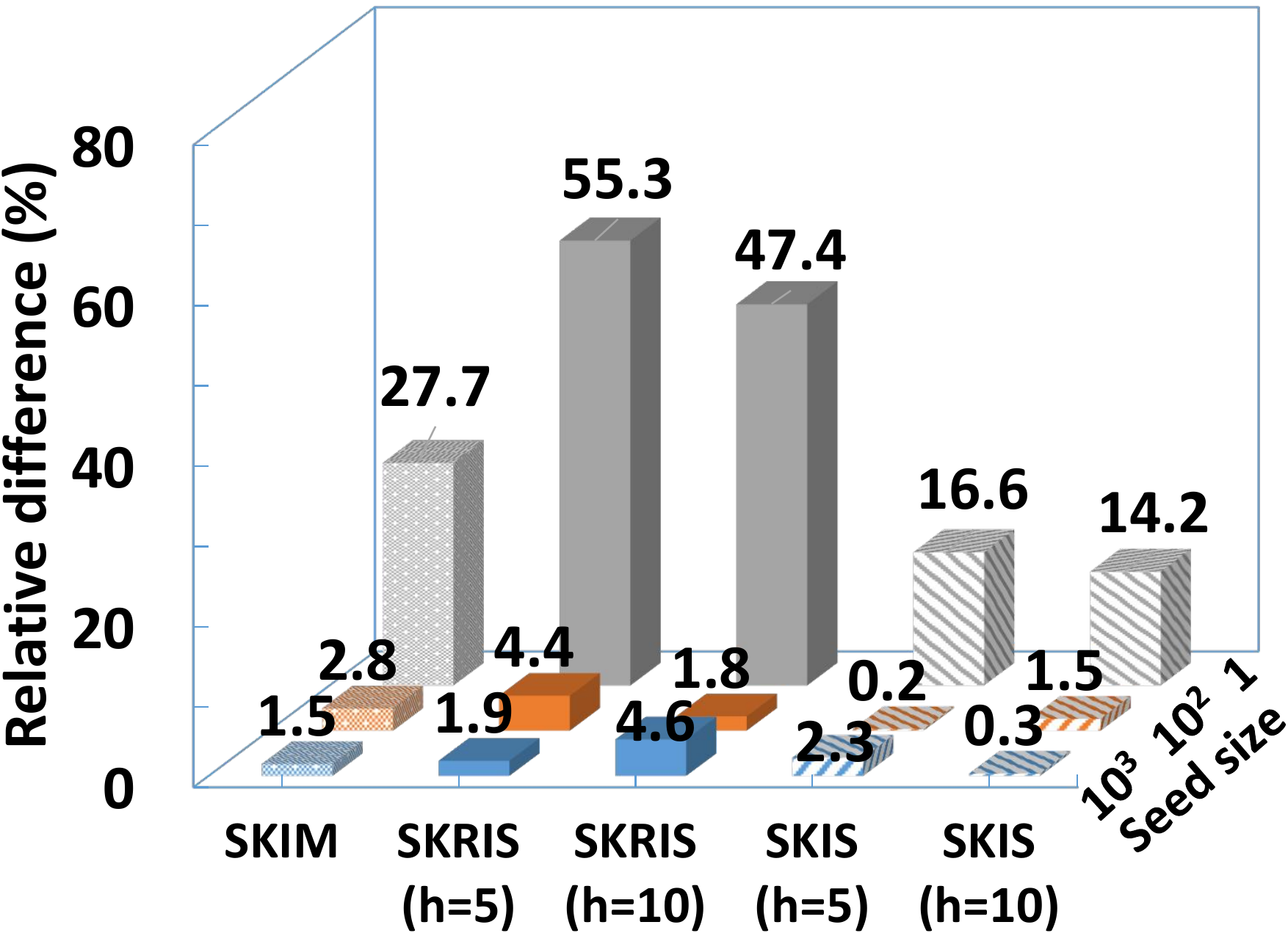}
	}
	\subfloat[\SKLRIS{}-\textsf{WC} model]{
		\includegraphics[width=0.22\linewidth]{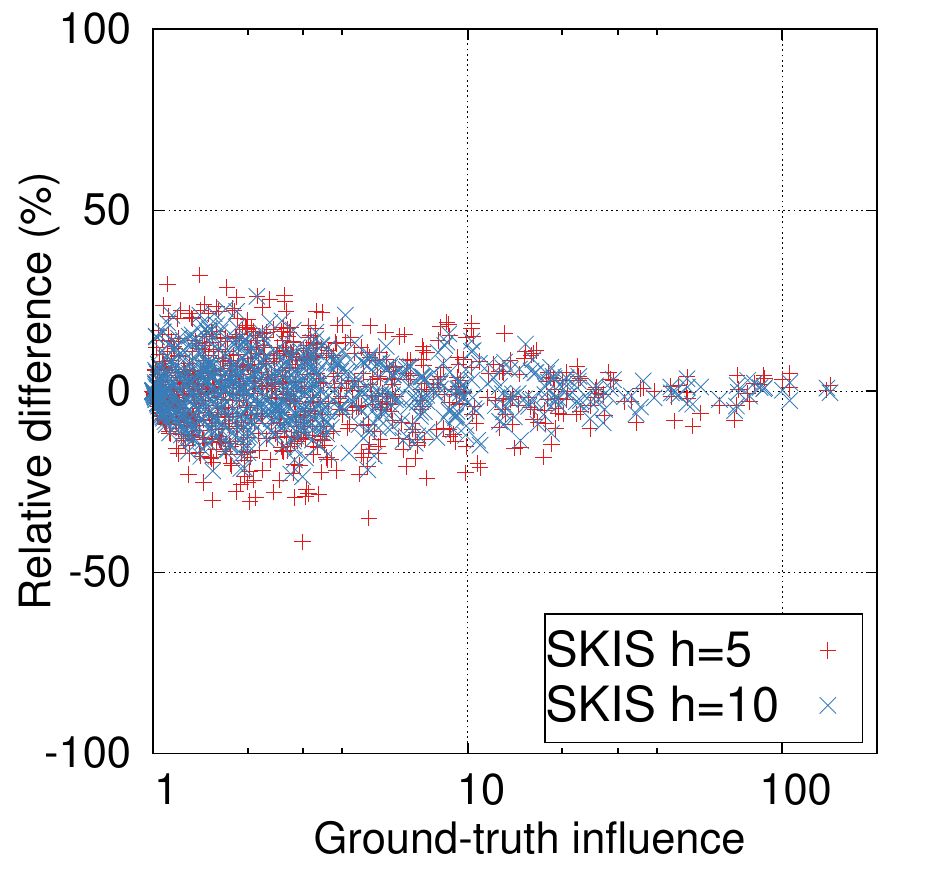}
	}
	\subfloat[\SKRIS{}-\textsf{WC} model]{
		\includegraphics[width=0.22\linewidth]{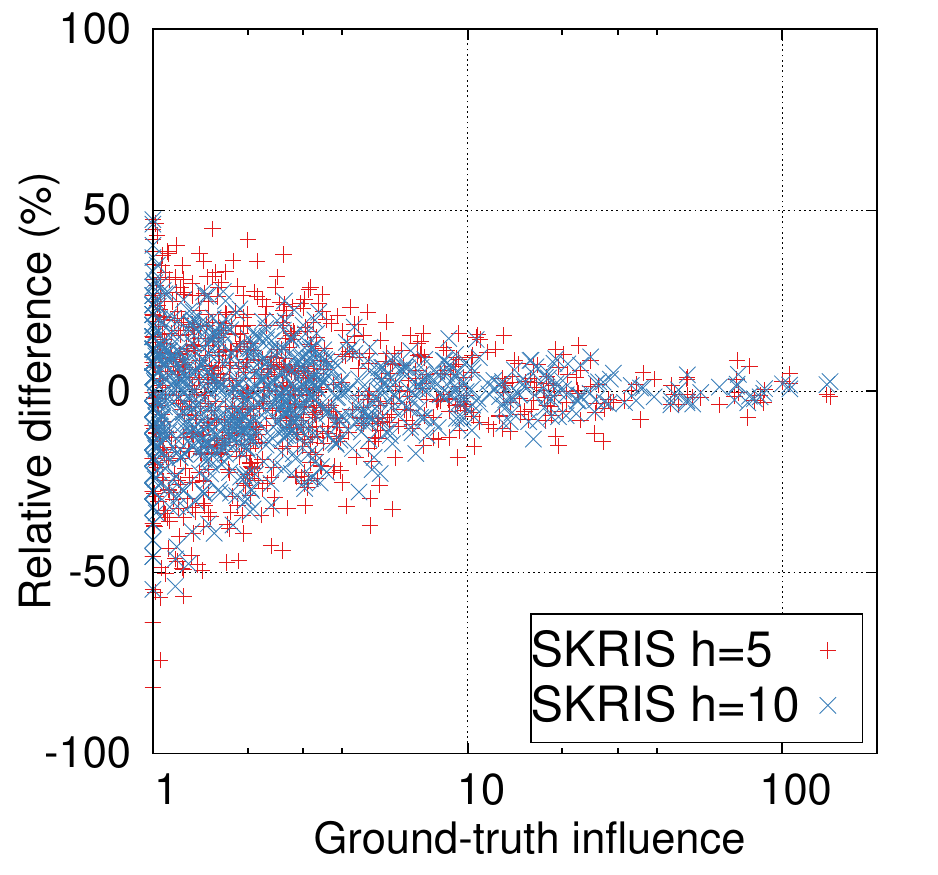}
	}
	\subfloat[\SKIM{}-\textsf{WC} model]{
		\includegraphics[width=0.22\linewidth]{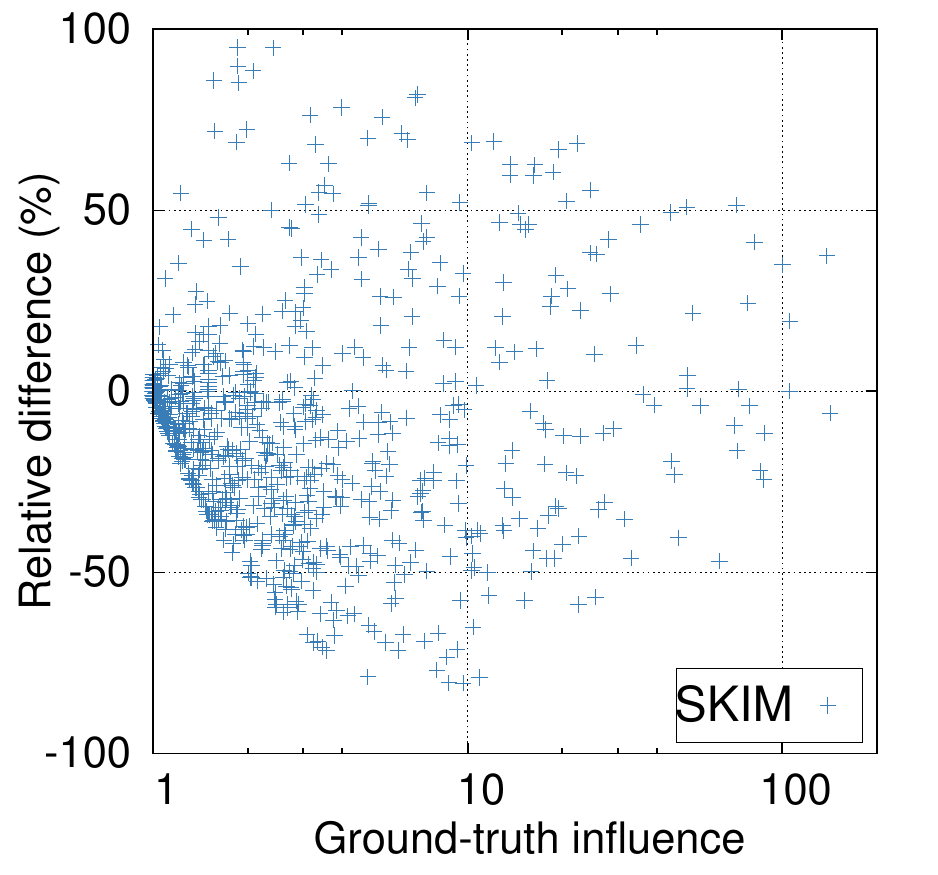}
	}
	\caption{a) Relative difference on Epinions under \textsf{TRI} model and b), c), d) error distributions under \textsf{WC} model with $|S| = 1$. \SKLRIS{} has the lowest relative errors which highly concentrates around 0 while \SKRIS{}'s and \SKIM{}'s errors widely spread out.}
	\label{fig:epi_err_wc}
	\vspace{-0.3in}
\end{figure*}

\begin{figure*}[!ht]
	\centering
	\subfloat[Epinions]{
		\includegraphics[width=0.24\linewidth]{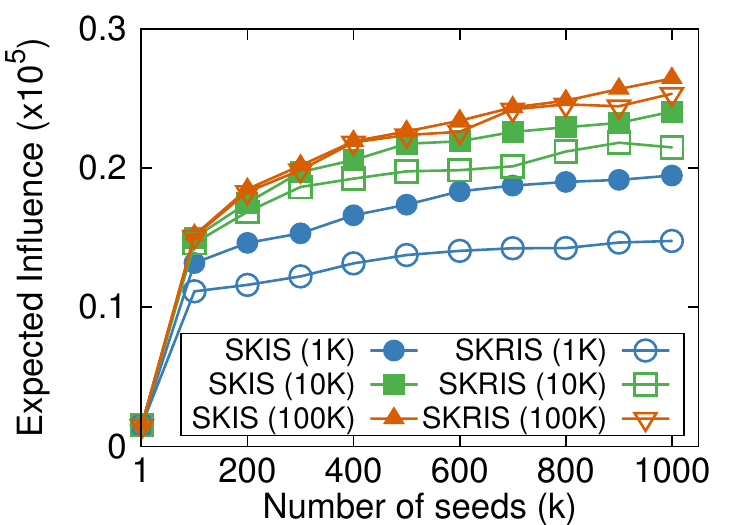}
	}
	\subfloat[NetPHY]{
		\includegraphics[width=0.24\linewidth]{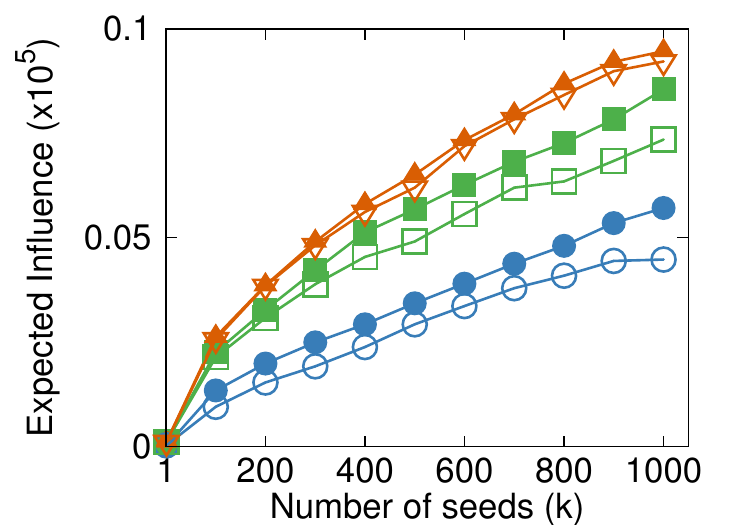}
	}
	\subfloat[DBLP]{
		\includegraphics[width=0.24\linewidth]{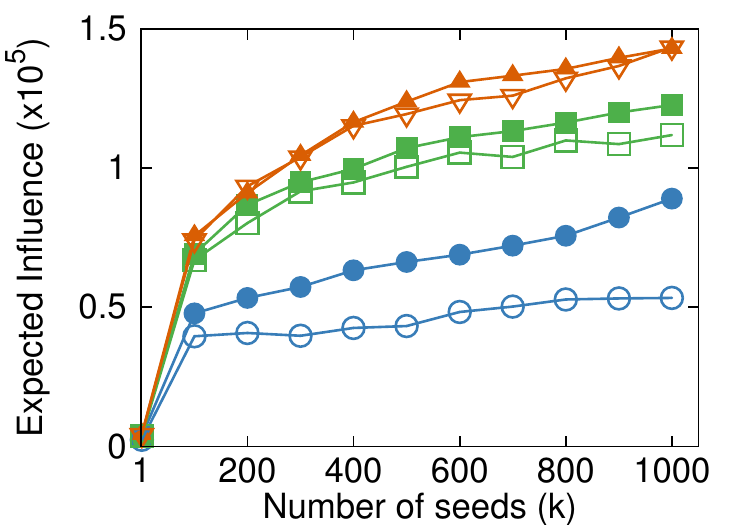}
	} 	
	\subfloat[Twitter]{
		\includegraphics[width=0.24\linewidth]{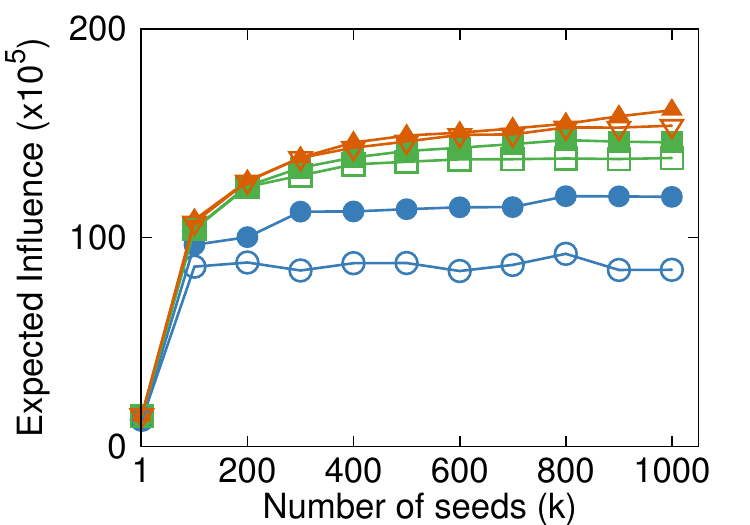}
	}
	\caption{Efficiency of \SKLRIS{} and \SKRIS{} sketches in finding the maximum seed sets. \SKLRIS{} sketch is up to 80\% more efficient.}
	\label{fig:im_inf_wc}
	\vspace{-0.25in}
\end{figure*}

\input{body/inf_est}


\setlength\tabcolsep{3pt}
\begin{table*}[!h] \small
	\centering
	\caption{Performance of \IM{} algorithms with $k=100$ (\textsf{dnf}: ``did not finish" within 6h, \textsf{mem}: ``out of memory"). }
	\begin{tabular}{cl rrrr r rrrr r rrrr r rrr}
		\toprule
		& & \multicolumn{4}{c}{\textbf{Running Time [s (or h)]}} && \multicolumn{4}{c}{\textbf{Total Memory [M (or G)]}} &&  \multicolumn{4}{c}{\textbf{Expected Influence (\%)}} && \multicolumn{3}{c}{\textbf{\#Samples [$\times 10^3$] }} \\
		\cmidrule{3-6}  \cmidrule{8-11} \cmidrule{13-16} \cmidrule{18-20}
		\multirow{2}{0.2in}{\textbf{}}& \multirow{2}{0.25in}{\textbf{Nets}} & \textsf{IMM} & \textsf{PMC}& \DPIMA{} & \DPIMA{}  && \textsf{IMM} & \textsf{PMC} & \DPIMA{} & \DPIMA{}  && \textsf{IMM} & \textsf{PMC}&  \DPIMA{} & \DPIMA{}  && \textsf{IMM} & \DPIMA{} & \DPIMA{} \\
		& && &  & +\textsf{SKIS}  &&&  &   & +\textsf{SKIS} &&&&   & +\textsf{SKIS}& &    &  & +\textsf{SKIS}   \\
		
		\midrule			
		\multirow{6}{*}{\textsf{WC}}
		& PHY  & 0.1  & 3.1 & \textbf{0.0} & \textbf{0.0}   && 31 & 86  & 26 &  \textbf{9}   && 6.64 & 6.7& 5.33 & 5.34   && 103.3  & 8.9 & \textbf{3.8}  \\
		& Epin. & 0.2  & 10.5  & \textbf{0.0} & \textbf{0.0}  && 39 & 130 & 34 &  \textbf{17}  && 19.4 & 19.8 & 17.9 & 16.6 && 39.8  & 4.48 & \textbf{0.9}  \\
		& DBLP  & 1.1 & 137.4  & \textbf{0.1} & \textbf{0.1} && 162 & \textbf{60} & 136 &  113   && 10.8 &11.2 & 9.3 & 8.5 && 93.0 & 5.4 & \textbf{2.6}\\
		& Orkut  & 24.1  & 1.4h  & 2.6 & \textbf{0.9}  && 4G & 6G  & \textbf{2G} &  \textbf{2G}   && 6.7  & 8.7 & 5.7  & 5.1 && 174.4 & 11.52 & \textbf{2.6}  \\
		& Twit.  & 67.3 & \textsf{mem} & 5.5 & 6.3  && 30G & \textsf{mem}  & 17G & \textbf{16G}  && 25.80  & \textsf{mem}  & 24.1 & 21.0 && 54.0 & 18.0 & \textbf{0.8} \\	
		& Frien.  & \textsf{dnf} & \textsf{mem}  & 78.3  & \textbf{43.6} && \textsf{dnf}  & \textsf{mem} & \textbf{35G} &  36G  && \textsf{dnf} & \textsf{mem}  & 0.35 & 0.35  &&  \textsf{mem} & 215.0 & \textbf{102.4}\\
		\midrule 
		\multirow{6}{*}{\textsf{TRI}} 
		& PHY  & 0.2  & 1.5  & \textbf{0.0} & \textbf{0.0} && 50 & 61  & 30 & \textbf{9}   && 1.77 & 1.73 & 1.4 & 1.5  && 370.1 & 35.8 & \textbf{3.8}  \\
		& Epin. & 13.9  & 6.9   & 2.0 & \textbf{0.6}   &&  483 & 40  & 72 & \textbf{33} && 5.7  & 5.9 & 5.47 & 5.46 && 123.0 & 8.9 & \textbf{0.5}  \\
		& DBLP   & 3.2 & 20.1  & 0.3 & \textbf{0.2} && 389  & \textbf{54} & 191 & 118   && 0.32 & 0.31 & 0.28 & 0.24  && 3171.0  & 348.2 & \textbf{20.5} \\
		& Orkut  &  \textsf{dnf} & 0.3h & 1.3h & \textbf{0.2h}   && \textsf{dnf} & 16G  & 28G & \textbf{11G}   &&  \textsf{dnf} & 67.3 & 67.9 & 67.8  && \textsf{dnf} & 1.4 & \textbf{0.3}  \\
		& Twit. & \textsf{dnf}  & \textsf{mem}  & 5.2h &  \textbf{0.6h}  && \textsf{dnf} & \textsf{mem} & 100G & \textbf{28G} && \textsf{dnf} & \textsf{mem}  & 24.2 & 24.4   && \textsf{dnf} & 3.4 &  \textbf{0.4}  \\	
		& Frien. & \textsf{dnf}  & \textsf{mem}  & \textsf{mem} & \textbf{3.1h} && \textsf{dnf} & \textsf{mem}  & \textsf{mem} &  \textbf{99G}  && \textsf{dnf}  & \textsf{mem}   & \textsf{mem} & 40.1  &&  \textsf{dnf}  & \textsf{mem} &  \textbf{0.2}  \\    		
		\bottomrule    		
	\end{tabular}
	\label{tab:compare_im}
	\vspace{-0.2in}
\end{table*}

\subsection{Influence Maximization}

This subsection illustrates the advantage of \ROS{} sketch in finding the seed set with maximum influence. The results show that \ROS{} samples drastically speed up the computation time. \DPIMA{}+\SKLRIS{} is the first to handle billion-scale networks on the challenging \textsf{TRI} edge model. We limit the running time for algorithms to 6 hours and put ``\textsf{dnf}'' if they cannot finish.
\subsubsection{Identifiability of the Maximum Seed Sets}
We compare the ability of the new \ROS{} with the traditional \RIS{} sampling in terms of identifying the seed set with maximum influence. We fix the number of samples generated to be in the set $\{1000,10000, 100000\}$ and then apply the \textsf{Greedy} algorithm to find solutions. We recompute the influence of returned seed sets using Monte-Carlo method with precision parameters $\epsilon = 0.005, \delta = 1/n$. The results is presented in Figure~\ref{fig:im_inf_wc}.

From Figure~\ref{fig:im_inf_wc}, we observe a recurrent consistency that \ROS{} samples return a better solution than \RIS{} over all the networks, $k$ values and number of samples. Particularly, the solutions provided by \ROS{} achieve up to $80\%$ better than those returned by \RIS{}. When more samples are used, the gap gets smaller.


\subsubsection{Efficiency of \SKLRIS{} on \IM{} problem}

Table~\ref{tab:compare_im} presents the results of \DPIMA{}-\textsf{SKIS}, \DPIMA{}, \IMM{} and \textsf{PMC} in terms of running time, memory consumption and samples generated.

\textbf{Running Time.} From Table~\ref{tab:compare_im}, the combination \DPIMA{}+\SKLRIS{} outperforms the rest by significant margins on all datasets and edge models. \DPIMA{}-\SKLRIS{} is up to 10x faster than the original \DPIMA. \DPIMA{}+\SKLRIS{} is the first and only algorithm that can run on the largest network on \textsf{TRI} model. 


\begin{figure}[!ht]
	\vspace{-0.2in}
	\centering
	\subfloat[Epinions]{
		\includegraphics[width=0.47\linewidth]{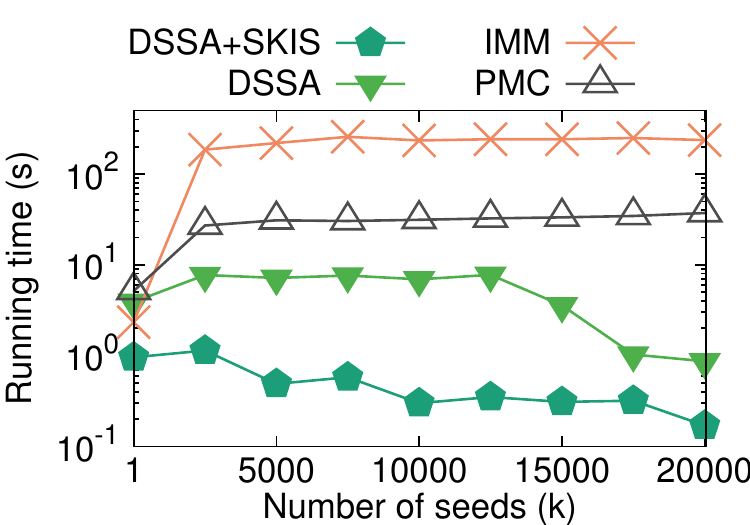}
	} 
	\subfloat[DBLP]{
		\includegraphics[width=0.47\linewidth]{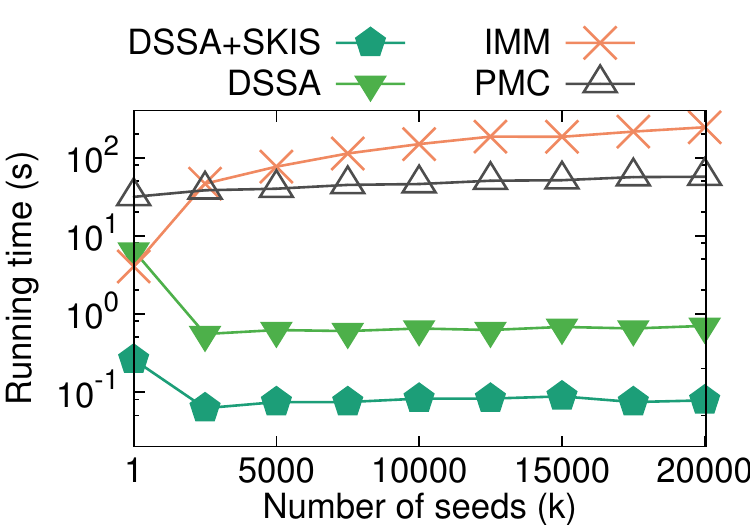}
	}
	\caption{Running time of algorithms under the IC model.}
	\label{fig:im_time_mod}
	\vspace{-0.35in}
\end{figure}
\begin{figure}[!ht]
	\centering
	\subfloat[Epinions]{
		\includegraphics[width=0.47\linewidth]{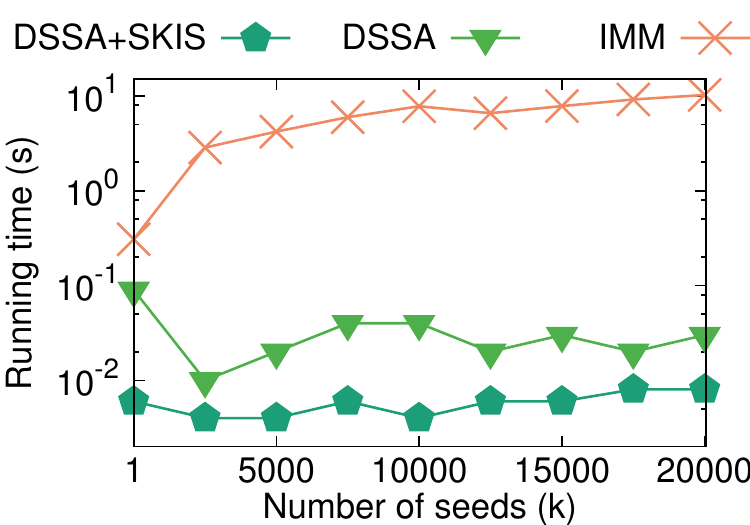}
	} 
	\subfloat[DBLP]{
		\includegraphics[width=0.47\linewidth]{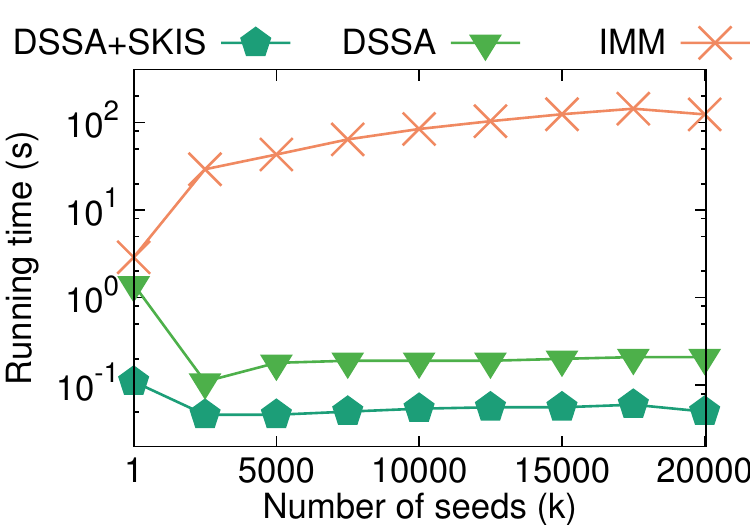}
	}
	\caption{Running time of algorithms under the LT model.}
	\label{fig:im_time_mod_lt}
	\vspace{-0.35in}
\end{figure}

Figure~\ref{fig:im_time_mod} compares the running time of all \IM{} algorithms across a wide range of budget $k = 1..20000$ under IC and \textsf{TRI} edge weight model. \DPIMA{}+\SKLRIS{} always maintains significant performance gaps to the other algorithms, e.g. 10x faster than \DPIMA{} or 1000x faster than \IMM{} and \textsf{PMC}.


\textbf{Number of Samples and Memory Usage.} On the same line with the running time, the memory usage and number of samples generated by \DPIMA{}+\SKLRIS{} are much less than those required by the other algorithms. The number of samples generated by \DPIMA{}+\SKLRIS{} is up to more 10x smaller than \DPIMA{} on \textsf{TRI} model, 100x less than \IMM{}. Since the memory for storing the graph is counted into the total memory, the memory saved by \DPIMA{}+\SKLRIS{} is only several times smaller than those of \DPIMA{} and \IMM{}. \textsf{PMC} exceptionally requires huge memory and is unable to run on two large networks.

\renewcommand{\arraystretch}{0.8}

\textbf{Experiments on the Linear Threshold (LT) model.} We carry another set of experiments on the LT model with multiple budget $k$. Since in LT, the total weights of incoming edge to every node are bounded by 1, for each node, we first normalized the weights of incoming edges and then multiply them with a random number uniformly generated in $[0,1]$.

The results are illustrated in Figure~\ref{fig:im_time_mod_lt}. Similar observations to the IC are seen in the LT model that \DPIMA+\SKLRIS{} runs faster than the others by orders of magnitude.




Overall, \DPIMA{}+\SKLRIS{} reveals significant improvements over the state-of-the-art algorithms on influence maximization. As a result, \DPIMA{}+\SKLRIS{} is the only algorithm that can handle the largest networks under different models.

%% file: body/inf_est.tex
\subsection{Influence Estimation}

We show that \SKLRIS{} sketch consumes much less time and memory space while consistently obtaining better solution quality, i.e. very small errors, than both \RIS{} and \SKIM{}. 


\subsubsection{Solution Quality}

Table~\ref{tbl:diff2_wc_all} and Figure~\ref{fig:epi_err_wc} present the relative estimation errors of all three sketches.

The solution quality of \SKLRIS{} is consistently better than \SKRIS{} and \SKIM{} across all the networks and edge models. As shown in Table~\ref{tbl:diff2_wc_all}, the errors of \SKLRIS{} are 110\% and 400\% smaller than those of \SKRIS{} with $k = 1$ while being as good as or better than \SKRIS{} for $k = 100, 1000$. On the other hand, \SKIM{} shows the largest estimation errors in most of the cases. Particularly, \SKIM{}'s error is more than 60 times higher than \SKLRIS{} and \SKRIS{} on Twitter when $|S| = 100$. Similar results are observed under \textsf{TRI} model. Exceptionally, on Twitter and Friendster, the relative difference of \SKRIS{} is slightly smaller than \SKLRIS{} with $h=5$ but larger on $h=10$. In \textsf{TRI} model, estimating a random seed on large network as Twitter produces higher errors since we have insufficient number of samples.

Figures~\ref{fig:epi_err_wc}b, c, and d draw the error distributions of sketches for estimating the influences of random seeds. Here, we generate 1000 uniformly random nodes and consider each node to be a seed set. We observe that \SKLRIS{}'s errors are highly concentrated around 0\% even when the influences are small while errors of \SKRIS{} and \SKIM{} spread out widely. \SKRIS{} reveals extremely high errors for small influence estimation, e.g. up to $80\%$. The error distribution of \SKIM{} is the most widely behaved, i.e. having high errors at every influence level. Under \textsf{TRI} model (Figure~\ref{fig:epi_err_wc}a), \SKLRIS{} also consistently provides significantly smaller estimation errors than \RIS{} and \SKIM{}.


\subsubsection{Performance}

We report indexing time and memory of different sketches in Table~\ref{tbl:sk_cstr_all}.

\textbf{Indexing Time.} \SKLRIS{} and \SKRIS{} use roughly the same amount of time for build the sketches while \SKIM{} is much slower than \SKLRIS{} and \SKRIS{} and failed to process large networks in both edge models. On larger networks, \SKLRIS{} is slightly faster than \SKRIS{}. \SKIM{} markedly spends up to 5 hours to build sketch for Twitter on \textsf{WC} model while \SKLRIS{}, or \SKRIS{} spends only 1 hour or less on this network. 

\textbf{Index Memory.} In terms of memory, the same observations are seen as with indexing time essentially because larger sketches require more time to construct. In all the experiments, \SKLRIS{} consumes the same or less amount of memory with \SKRIS{}. \SKIM{} generally uses more memory than \SKLRIS{} and \SKRIS{}.






In summary, \SKLRIS{} consistently achieves better solution quality than both \SKRIS{} and \SKIM{} on all the networks, edge models and seed set sizes while consuming the same or less time/memory. The errors of \SKLRIS{} is highly concentrated around 0. In contrast, \SKRIS{} is only good for estimating high influence while incurring significant errors for small ranges.


%% file: body/appendix.tex
\subsection{Proof of Lemma~\ref{lem:multi_set_rois}}

	Given a stochastic graph $\G$, recall that $\Omega_\G$ is the space of all possible sample graphs $g \sim \G$ and $\Pr[g]$ is the probability that $g$ is realized from $\G$. In a sample graph $g \in \Omega_\G$, $\eta_g(S,v) = 1$ if $v$ is reachable from $S$ in $g$. Consider the graph sample space $\Omega_\G$, based on a node $v \in V \backslash S$, we can divide $\Omega_\G$ into two partitions: 1) $\Omega^\emptyset_\G(v)$ contains those samples $g$ in which $v$ has \textit{no incoming live-edges}; and 2) $\bar \Omega^\emptyset_\G(v) = \Omega_{\G} \backslash \Omega^\emptyset_\G$. We start from the definition of influence spread as follows,
	\begin{align}
		\I(S) & = \sum_{v \in V} \sum_{g \in \Omega_\G} \eta_g(S,v) \Pr[g] \nonumber \\
		& = \sum_{v \in V} \Big (\sum_{g \in \Omega^\emptyset_\G(v)} \eta_g(S,v) \Pr[g] + \sum_{g \in \bar \Omega^\emptyset_\G(v)} \eta_g(S,v) \Pr[g]\Big ). \nonumber
	\end{align}
	
	In each $g \in \Omega^\emptyset_\G(v)$, the node $v$ does not have any incoming nodes, thus, $\eta_g(S,v) = 1$ only if $v \in S$. Thus, we have that $\sum_{v \in V} \sum_{g \in \Omega^\emptyset_\G(v)} \eta_g(S,v) \Pr[g] = \sum_{v \in S} \sum_{g \in \Omega^\emptyset_\G(v)} \Pr[g]$. Furthermore, the probability of a sample graph which has no incoming live-edge to $v$ is $ \sum_{g \in \Omega^\emptyset_\G(v)} \Pr[g] = 1-\gamma_v$. Combine with the above equiation of $\I(S)$, we obtain,
	\begin{align}
		\label{eq:lem6_eq1}
		\I(S) = \sum_{v\in S} (1-\gamma_v) + \sum_{v \in V} \sum_{g \in \bar \Omega^\emptyset_\G(v)} \eta_g(S,v) \Pr[g \in \Omega_\G].
	\end{align}
	
	Since our \ROS{} sketching algorithm only generates samples corresponding to sample graphs from the set $\bar \Omega^\emptyset_\G(v)$, we define $\bar \Omega^\emptyset_\G(v)$ to be a graph sample space in which the sample graph $\bar g \in \bar \Omega^\emptyset_\G(v)$ has a probability $\Pr[\bar g \in \bar \Omega^\emptyset_\G(v)] = \frac{\Pr[\bar g \in \Omega_\G]}{\gamma_v}$ of being realized (since $\sum_{\bar g \in \bar \Omega^\emptyset_\G(v)}\Pr[\bar g \in \Omega_\G] = \gamma_v$ is the normalizing factor to fulfill a probability distribution of a sample space). Then, Eq.~\ref{eq:lem6_eq1} is rewritten as follows,
	\begin{align}
		\I(S) & = \sum_{v \in V} \sum_{g \in \bar \Omega^\emptyset_\G(v)} \eta_g(S,v) \frac{\Pr[g \in \Omega_\G]}{\gamma_v} \gamma_v + \sum_{v\in S} (1-\gamma_v) \nonumber \\
		& = \sum_{v \in V} \sum_{\bar g \in \bar \Omega^\emptyset_\G(v)} \eta_{\bar g}(S,v) \Pr[\bar g \in \bar \Omega^\emptyset_\G(v)] \gamma_v + \sum_{v\in S} (1-\gamma_v) \nonumber
	\end{align}
	
	Now, from the node $v$ in a sample graph $\bar g \in \bar \Omega^\emptyset_\G(v)$, we have a \ROS{} sketch $R_j(\bar g, v)$ starting from $v$ and containing all the nodes that can reach $v$ in $\bar g$. Thus, $\eta_{\bar g}(S,v) = \textbf{1}_{R_j(\bar g,v) \cap S \neq \emptyset}$ where $\textbf{1}_{x}$ is an indicator function returning 1 iff $x \neq 0$. Then,
	\begin{align}	
		& \sum_{\bar g \in \bar \Omega^\emptyset_\G(v)} \eta_{\bar g}(S,v) \Pr[\bar g \in \bar \Omega^\emptyset_\G(v)] \nonumber \\
		& = \sum_{\bar g \in \bar \Omega^\emptyset_\G(v)} \textbf{1}_{R_j(\bar g,v) \cap S \neq \emptyset} \Pr[\bar g \in \bar \Omega^\emptyset_\G(v)] = \Pr[R_j(v) \cap S \neq \emptyset] \nonumber
	\end{align}
	
	\noindent where $R_j(v)$ is a random \ROS{} sketch with $\src(R_j(v)) = v$.	Plugging this back into the computation of $\I(S)$ gives,
	\begin{align}
		& \I(S) = \sum_{v \in V} \Pr[R_j(v) \cap S \neq \emptyset] \gamma_v + \sum_{v\in S} (1-\gamma_v) \nonumber \\
		& = \sum_{v \in V} \Pr[R_j(v) \cap S \neq \emptyset] \frac{\gamma_v}{\Gamma} \Gamma + \sum_{v\in S} (1-\gamma_v) \nonumber \\
		& = \sum_{v \in V} \Pr[R_j(v) \cap S \neq \emptyset] \Pr[\src(R_j) = v]\Gamma + \sum_{v\in S} (1-\gamma_v) \nonumber \\
		& = \Pr[R_j \cap S \neq \emptyset] \cdot \Gamma + \sum_{v\in S} (1-\gamma_v)
	\end{align}
	
	\noindent That completes the proof.

\subsection{Proof of Lemma~\ref{lem:zvar}}

	From the basic properties of variance, we have,
	\begin{align}
		\Var[Z_j(S)] & = \Var[\frac{X_j(S) \cdot \Gamma + \sum_{v \in S}(1-\gamma_v)}{n}] \nonumber \\
		& = \frac{\Gamma^2}{n^2} \Var[X_j(S)] \nonumber
	\end{align}
	
	Since $X_j(S)$ is a Bernoulli random variable with its mean value $\E[X_j(S)] =  \frac{\I(S)-\sum_{v\in S}(1-\gamma_v)}{\Gamma}$, the variance $\Var[X_j(S)]$ is computed as follows,
	\begin{align}
		& \Var[X_j(S)] \nonumber \\
		& = \frac{\I(S)-\sum_{v\in S}(1-\gamma_v)}{\Gamma} (1 - \frac{\I(S)-\sum_{v\in S}(1-\gamma_v)}{\Gamma}) \nonumber \\
		& = \frac{\I(S)}{\Gamma} - \frac{\I^2(S)}{\Gamma^2} - \frac{\sum_{v \in S}(1-\gamma_v)}{\Gamma^2}(\Gamma + \sum_{v \in S}(1-\gamma_v) - 2\I(S)) \nonumber
	\end{align}
	
	Put this back into the variance of $Z_j(S)$ proves the lemma.

\subsection{Proof of Lemma~\ref{lem:var_z}}
Since $Z_j(S)$ takes values of either $\frac{\sum_{v \in S}(1-\gamma_v)}{n}$ or $\frac{\Gamma + \sum_{v \in S}(1-\gamma_v)}{n}$ and the mean value $\E[Z_j(S)] = \frac{\I(S)}{n}$, i.e. $\frac{\sum_{v \in S}(1-\gamma_v)}{n} \leq \frac{\I(S)}{n} \leq \frac{\Gamma + \sum_{v \in S}(1-\gamma_v)}{n}$. The variance of $Z_j(S)$ is computed as follows,
\begin{align}
	\Var&[Z_j(S)] \nonumber \\
	& = \Big( \frac{\I(S)}{n} - \frac{\sum_{v \in S}(1-\gamma_v)}{n} \Big) \Big( \frac{\Gamma + \sum_{v \in S}(1-\gamma_v)}{n} - \frac{\I(S)}{n} \Big)\nonumber \\
	& \leq \frac{\I(S)}{n} \Big( \frac{\Gamma + \sum_{v \in S}(1-\gamma_v)}{n} - \frac{\sum_{v \in S}(1-\gamma_v)}{n} \Big) \nonumber \\
	& = \frac{\I(S)}{n} \frac{\Gamma}{n}
\end{align}

\subsection{Proof of Lemma~\ref{lem:bound}}
Lemma~2 in \cite{Tang15} states that:
\begin{Lemma}
	Let $M_1, M_2, \dots$ be a martingale, such that $|M_1| \leq a$, $|M_j - M_{j-1}| \leq a$ for any $j \in [2, T]$, and
	\begin{align}
	\Var[M_1] + \sum_{j = 2}^{T} \Var[M_j | M_1, M_2, \dots, M_{j-1}] \leq b,
	\end{align}
	where $\Var[.]$ denotes the variances of a random variable. Then, for any $\lambda > 0$,
	\begin{align}
	\label{eq:mar_cher}
	\Pr[M_T - \E[M_T] \geq \lambda] \leq \exp\Big( - \frac{\lambda^2}{\frac{2}{3}a\lambda + 2b} \Big)
	\end{align}
\end{Lemma}

Note that uniform random variables are also a special type of martingale and the above lemma holds for random variable as well.
Let $p = \frac{\I(S)}{n}$. For \RIS{} samples, since 
\begin{itemize}
	\item $|M_1| \leq 1$, 
	\item $|M_j - M_{j-1}| \leq 1, \forall j \in [2,T]$,
	\item $\Var[M_1] + \sum_{j = 2}^{T}\Var[M_j | M_1, \dots, M_{j-1}] = \sum_{j = 1}^{i} \Var[Y_j(S)] = T p(1-p) \leq T p$,
\end{itemize}
applying Eq.~\ref{eq:mar_cher} for $\lambda = \epsilon Tp$ gives the following Chernoff's bounds,
\begin{align}
\label{eq:mar_cher_ris_1}
&\Pr\Big[\sum_{j = 1}^{T} X_j(S) - Tp \geq \epsilon Tp\Big] \leq \exp\Big( -\frac{\epsilon^2}{2+\frac{2}{3}\epsilon} Tp \Big), \\
&\text{and,} \nonumber \\
\label{eq:mar_cher_ris_2}
&\Pr\Big[\sum_{j = 1}^{T} X_j(S) - Tp \leq -\epsilon Tp\Big] \leq \exp\Big( -\frac{\epsilon^2}{2} Tp \Big).
\end{align}

However, for \ROS{} samples in \SKLRIS{} sketch, the corresponding random variables $Z_j(S)$ replace $Y_j(S)$ and have the following properties:
\begin{itemize}
	\item $|M_1| \leq \frac{\Gamma + \sum_{v \in S}(1-\gamma_v)}{n} \leq 1$, 
	\item $|M_j - M_{j-1}| \leq 1, \forall j \in [2,T]$,
	\item The sum of variances:
	\begin{align}
	&\Var[M_1] + \sum_{j = 2}^{T}\Var[M_j | M_1, \dots, M_{j-1}] \nonumber \\
	&= \sum_{j = 1}^{i} \Var[Z_j(S)] = T p \frac{\Gamma}{n}
	\end{align}
\end{itemize}
Thus, applying the general bound in Eq.~\ref{eq:mar_cher} gives,
\begin{align}
&\Pr\Big[\sum_{j = 1}^{T} X_j(S) - Tp \geq \epsilon Tp\Big] \leq \exp\Big( -\frac{\epsilon^2}{2\boldsymbol{\frac{\Gamma}{n}}+\frac{2}{3}\epsilon} Tp \Big), \\
&\text{and,} \nonumber \\
&\Pr\Big[\sum_{j = 1}^{T} X_j(S) - Tp \leq -\epsilon Tp\Big] \leq \exp\Big( -\frac{\epsilon^2}{2\boldsymbol{\frac{\Gamma}{n}}} Tp \Big).
\end{align}

Note the factor $\frac{\Gamma}{n}$ is added in the denominator of the terms in the $\exp(.)$ function. Since $2\frac{\Gamma}{n}$ dominates $\frac{2}{3}\epsilon$, the concentration bounds for $Z_j(S)$ for \SKLRIS{} are tighter than those of $Y_j(S)$ for \RIS{} given in Eqs.~\ref{eq:mar_cher_ris_1} and \ref{eq:mar_cher_ris_2}.

\subsection{Proof of Lemma~\ref{lem:sub}}

	Since the function $\hat \I_\R(S)$ contains two additive terms, it is sufficient to show that each of them is monotone and submodular. The second term $\sum_{v \in S}(1-\gamma_v)$ is a linear function and thus, it is monotone and submodular. For the first additive term, we see that $\frac{\Gamma}{|\R| \cdot n}$ is a constant and only need to show that $\Cov_\R(S)$ is monotone and submodular. Given the collection of \ROS{} samples $\R$ in which $R_j \in \R$ is a list of nodes, the function $\Cov_\R(S)$ is just the count of \ROS{} samples that intersect with the set $S$. In other words, it is equivalent to a covering function in a set system where \ROS{} samples are elements and nodes are sets. A set covers an element if the corresponding node is contained in the corresponding \ROS{} sample. It is well known that any covering function is monotone and submodular \cite{Vazirani01} and thus, the $\Cov_\R(S)$ has the same properties.

\subsection{Improvements of \PIMA/\DPIMA{} using \SKLRIS{} sketch}
Recall that the original Stop-and-Stare strategy in \cite{Nguyen163} uses two independent sets of \RIS{} samples, called $\R$ and $\R^c$. The greedy algorithm is applied on the first set $\R$ to find a candidate set $\hat S_k$ along with an estimate $\hat \I_\R(\hat S_k)$ and the second set $\R^c$ is used to reestimate the influence of $\hat S_k$ by $\hat \I_{\R^c}(\hat S_k)$. Now, \PIMA{} and \DPIMA{} have different ways to check the solution quality.

\textbf{\PIMA{}.} It assumes a set of fixed precision parameters $\epsilon_1, \epsilon_2, \epsilon_3$ such that $\frac{\epsilon_1 + \epsilon_2 + \epsilon_1 \epsilon_2 + \epsilon_3}{(1+\epsilon_1)(1+\epsilon_2)} (1-1/e) \leq \epsilon$. The algorithm stops when two conditions are met:
\begin{itemize}
	\item[1)] $\Cov_\R(\hat S_k) \geq \Lambda_1$ where $ \Lambda_1 = O(\log\frac{\delta}{t_{max}}\epsilon_3^{-2})$ and $t_{max}$ is a precomputed number depending on the size of the input graph $\G$.
	\item[2)] $\hat \I_\R(\hat S_k) \leq (1+\epsilon_1) \hat \I_{\R^c}(\hat S_k)$.
\end{itemize}

$\Rightarrow$ \textit{Improvements using \SKLRIS{}:} Replacing \RIS{} samples by \ROS{} samples to build $\R$ and $\R^c$ helps in both stopping conditions:
\begin{itemize}
	\item Reduce $\Lambda_1$ to $\Lambda_1 = O(\frac{\Gamma}{n}\log\frac{\delta}{t_{max}}\epsilon_3^{-2})$ using the tighter form of the Chernoff's bounds in Lemma~\ref{lem:bound}.
	\item Since \ROS{} samples have better influence estimation accuracy, $\hat \I_\R(\hat S_k)$ and $\hat \I_{\R^c}(\hat S_k)$ are closer to the true influence $\I(\hat S_k)$. Thus, the second condition is met earlier than using \RIS{} samples.
\end{itemize}

\textbf{\DPIMA.} Instead of assuming precision parameters, \DPIMA{} dynamically compute the error bounds $\epsilon_1, \epsilon_2$ and $\epsilon_3$ as follows:
\begin{itemize}
	\item $\epsilon_1 = \frac{\hat \I_{\R}(\hat S_k)}{\hat \I_{\R^c}(\hat S_k)} - 1$.
	\item $\epsilon_2 = \epsilon \sqrt{\frac{n(1+\epsilon)}{2^{t-1}\hat \I_{\R^c}(\hat S_k)}}$.
	\item $\epsilon_3 = \epsilon \sqrt{\frac{n(1+\epsilon)(1-1/e-\epsilon)}{(1+\epsilon/3)2^{t-1}\hat \I_{\R^c}(\hat S_k)}}$.
\end{itemize}
Here, $\epsilon_1$ measures the discrepancy of estimations using two different sketches $\R$ and $\R^c$ while $\epsilon_2$ and $\epsilon_3$ are the error bounds of estimating the influences of $\hat S_k$ and the optimal solution $S^*_k$ using the number of samples contained in $\R$ and $\R^c$. The algorithm stops when two conditions are met:
\begin{itemize}
	\item $\Cov_\R(\hat S_k) \geq \Lambda_2$ where $\Lambda_2 = O(\log\frac{\delta}{t_{max}} \epsilon^{-2})$.
	\item $(\epsilon_1 + \epsilon_2+\epsilon_1\epsilon_2)(1-1/e-\epsilon)+(1-1/e)\epsilon_3 \leq \epsilon$.
\end{itemize}
$\Rightarrow$ Improvement using \SKLRIS{}: Similarly to \PIMA{}, applying \SKLRIS{} helps in both stopping conditions:
\begin{itemize}
	\item Reduce $\Lambda_2$ to $\Lambda_2 = O(\frac{\Gamma}{n}\log\frac{\delta}{t_{max}} \epsilon^{-2})$.
	\item Reduce the value of $\epsilon_1, \epsilon_2$ and $\epsilon_3$ due to better influence estimations of $\hat \I_{\R}(\hat S_k)$ and $\hat \I_{\R^c}(\hat S_k)$ by \SKLRIS{} that leads to earlier satisfaction of the second condition.
\end{itemize}